\documentclass[12pt,preprintnumbers,superscriptaddress,amsmath,amssymb,nofootinbib]{revtex4}

\usepackage{url}
\usepackage{graphicx}
\usepackage{dcolumn}
\usepackage{bm}
\usepackage{amssymb}
\usepackage{amsmath}
\usepackage{epsfig}    
\usepackage{color}
\usepackage{slashed}
\usepackage[hypertex]{hyperref}

\begin{document} 

\title{H-COUP:\\ a program for one-loop corrected Higgs boson couplings\\
in non-minimal Higgs sectors }

\author{Shinya Kanemura}
\email{kanemu@phys.sci.osaka-u.ac.jp }
\affiliation{Department of Physics, Osaka University, Toyonaka, Osaka 560-0043, Japan}

\author{Mariko Kikuchi}
\email{marikokikuchi@hep1.phys.ntu.edu.tw}
\affiliation{Department of Physics, National Taiwan University, Taipei 10617, Taiwan}

\author{Kodai Sakurai}
\email{ksakurai@het.phys.sci.osaka-u.ac.jp}
\affiliation{Department of Physics, University of Toyama, \\3190 Gofuku, Toyama 930-8555, Japan}

\author{Kei Yagyu}
\email{yagyu@fi.infn.it}
\affiliation{INFN, Sezione di Firenze, and Department of Physics and Astronomy, University of Florence, Via
G. Sansone 1, 50019 Sesto Fiorentino, Italy}

\preprint{OU-HET 947}
\preprint{UT-HET 122}

\begin{abstract}

\begin{center}
{ \bf Abstract}
\end{center}
\vspace{-5mm}
We describe a numerical calculation tool {\tt H-COUP}\footnote{The webpage of {\tt H-COUP} is given in 
\url{http://www-het.phys.sci.osaka-u.ac.jp/~kanemu/HCOUP_HP1013/HCOUP_HP.html}, where one can download the set of source files of {\tt H-COUP\_1.0}.\\   } written in Fortran, which provides 
one-loop electroweak corrected vertices for the discovered Higgs boson $h(125)$ in various Higgs sectors. 
The renormalization is based on the improved on-shell scheme without gauge dependence. 
In the first version {\tt H-COUP\_1.0}, the following models are included, namely, 
the Higgs singlet model, four types (Type-I, Type-II, Type-X, Type-Y) of two Higgs doublet models with a softly-broken $Z_2$ symmetry and the inert doublet model. 
We first briefly introduce these models and then explain how to install and run this tool in an individual machine. 
A sample of numerical outputs is provided for user information. 

\end{abstract}
\maketitle

\newpage
\section{Introduction}

After the discovery of the Higgs boson $h(125)$ in 2012 at the LHC~\cite{LHC_Higgs_ATLAS, LHC_Higgs_CMS}, 
its property has been measured and it has turned out to be consistent with that of the Higgs boson in the Standard Model (SM)~\cite{Aad:2015zhl, Khachatryan:2016vau}. 
Nevertheless, there is an expectation that the SM is replaced by a new physics model at the TeV scale or higher by which phenomena beyond the SM such as 
neutrino oscillations, the existence of dark matter and baryon asymmetry of the Universe can be explained. 
If this is the case, some hint or evidence for new physics should be found by current or future experiments. 

In the SM, only one scalar isospin doublet field is introduced to break the electroweak gauge symmetry. 
This is just an assumption. There is a possibility for an extended Higgs sector with a specific multiplet structure. 
Actually, various new physics models beyond the SM predict characteristic non-minimal Higgs sectors.  
Therefore, by detailed studies of the Higgs sector we can narrow down models of new physics.  
Determining the structure of the Higgs sector is a top priority in the construction of the new physics theory. 

If the second Higgs boson is discovered at collider experiments, it must be direct evidence for non-minimal Higgs sectors. 
So far, there have been no report for the discovery of additional Higgs bosons. 
However, even without the direct discovery, we can find indirect evidence for non-minimal Higgs sectors, because 
the effect of non-minimality appears on the observables of $h(125)$ as deviations from the SM predictions. 
From the magnitude of the deviation one can obtain hints for the scale of the new physics such as the mass of the second Higgs boson. 
Furthermore, the pattern of the deviation in each observable like coupling constants strongly depends on details of the structure of the Higgs sector or scenarios of new physics.    
Therefore, precision measurements of the Higgs boson properties are essentially important to explore new physics beyond the SM.    

In the future, 
measurements of the mass, production cross sections, decay branching ratios and couplings and so on, are expected to be drastically improved by the LHC experiment, including its high luminosity option,  
and future lepton colliders such as International Linear Collider (ILC)~\cite{Barklow:2015tja, Tian:2016qlk}, 
Compact LInear Collider (CLIC)~\cite{CLIC:2016zwp} and $e^+e^-$ Future Circular Collider (FCC-ee)~\cite{Gomez-Ceballos:2013zzn}.  
At experiments in these future colliders, the Higgs boson couplings are expected to be measured with a percent or better accuracy.  
In order to compare to such precision measurements, 
the theory predictions should also be as accurate as possible with including higher order corrections in various extended Higgs sectors.  

\begin{table}[t]\begin{center}
\begin{tabular}{c|ccccccc|ccc}\hline\hline
              & $hf\bar{f}$ (QCD) & $hf\bar{f}$ (EW) & $hVV$ & $hhh$  \\\hline
SM            & \cite{hqq_NLO1,hqq_NLO2,hqq_NLO3,hqq_NLO4} &\cite{Fleischer,Kniehl_hff}&\cite{Fleischer,Kniehl_hzz,Kniehl_hww}& \cite{hhh-sm1,Kanemura:2004mg} \\\hline
MSSM          & \cite{Dabelstein:1995js, Jimenez:1995wf}  &\cite{Dabelstein:1995js, Jimenez:1995wf} & \cite{Chankowski:1992er} & \cite{Hollik:2001px, Dobado:2002jz, Carena:2001bg} \\\hline
THDMs          &    & \cite{Arhrib:2003ph, Kanemura:2014dja, Kanemura:2015mxa,Kanemura:2017wtm}      &  \cite{Kanemura:2004mg, LopezVal:2010vk, Kanemura:2015mxa, Altenkamp:2017ldc, Kanemura:2017wtm}  &\cite{Kanemura:2004mg, Kanemura:2015mxa, Kanemura:2017wtm}     \\\hline
HSMs           &    & \cite{Kanemura:2015fra,Kanemura:2017wtm}  & \cite{Kanemura:2015fra,Kanemura:2017wtm} & \cite{Kanemura:2016lkz, He:2016sqr,Kanemura:2017wtm} \\\hline 
IDM           &    & \cite{Kanemura:2016sos,Kanemura:2017wtm}          & \cite{Arhrib:2015hoa,Kanemura:2016sos,Kanemura:2017wtm} & \cite{Arhrib:2015hoa,Kanemura:2016sos,Kanemura:2017wtm} \\\hline\hline
\end{tabular}
\caption{Summary for studies on radiative corrections to the Higgs boson vertices at one-loop level. 
For the $hf\bar{f}$ vertex, we separately show the works for the one-loop QCD corrections (the first column) 
and electroweak/Higgs loop corrections (the second column). 
Higher QCD corrections to the $hf\bar{f}$ and the loop induced $hgg$ vertices are discussed in Sec.~\ref{sec:decay_rate}. 
}
\label{previous}
\end{center}
\end{table}

Radiative corrections to the Higgs boson vertices have been investigated in various Higgs sectors. 
In Table~\ref{previous}, we show studies on one-loop corrections to the $hf\bar{f}$, $hVV$ and $hhh$ vertices in the 
SM, two Higgs doublet models (THDMs) with a softly-broken $Z_2$ symmetry, 
Higgs singlet models (HSMs)\footnote{In fact, there are several versions of the HSM, e.g., a model with an $U(1)$ gauge symmetry, 
a discrete $Z_2$ symmetry and the most general case without imposing any additional symmetry. We here do not distinguish these variations.  }, and the inert doublet model (IDM). 

These radiative corrected vertices and their applications to the physical processes such as the decay rates and cross sections 
can be numerically evaluated by using several public numerical tools. 
{\tt HDECAY}~\cite{Djouadi:1997yw} and {\tt FeynHiggs}~\cite{Heinemeyer:1998yj,Hahn:2009zz} 
({\tt NMHDECAY}~\cite{Ellwanger:2004xm}) 
provides decay widths and decay branching ratios of Higgs bosons with electroweak and QCD corrections in the SM and the MSSM (next to MSSM).  
For non-supersymmetric (SUSY) models, {\tt 2HDMC}~\cite{Eriksson:2009ws} and {\tt sHDECAY}~\cite{Costa:2015llh}
give decay rates, total widths and branching fraction for Higgs bosons with QCD corrections in THDMs and HSMs, respectively. 
However, there is still no public program tool which evaluates observables of the Higgs boson with electroweak loop corrections in non-SUSY models with extended Higgs sectors. 

In this manual, we explain how to use a new calculation tool ``{\tt H-COUP\_1.0}'' written in a Fortran code to evaluate one-loop electroweak corrected 
Higgs boson vertices ($hWW$, $hZZ$, $ht\bar{t}$, $hb\bar{b}$, $hc\bar{c}$, $h\tau\tau$, $hhh$) and the loop induced decay rates ($h\to \gamma\gamma$, $h\to Z\gamma$, $h\to gg$) as well as 
the oblique electroweak $S$ and $T$ parameters in non-minimal Higgs sectors. 
{\tt H-COUP\_1.0} includes the HSM (a model with a real singlet scalar field), the four types (Type-I, Type-II, Type-X and Type-Y) 
of the THDM with a softly-broken $Z_2$ symmetry and the IDM in addition to the SM. 
The one-loop corrections are evaluated by the improved on-shell renormalization scheme defined in Ref.~\cite{Kanemura:2017wtm}, in which 
gauge dependence appearing in mixings among scalar fields is removed by the pinch technique~\cite{Yamada:2001px, Espinosa:2002cd}. 
The parameter space of each model is constrained by various theoretical and experimental bounds. 
With the future precision data, one can fingerprint the various Higgs boson couplings by comparing the future data with precise theoretical predictions calculated by using {\tt H-COUP\_1.0}, 
and identify the Higgs sector as discussed in Refs.~\cite{Kanemura:2014bqa, Kakizaki:2013eba, Kanemura:2014kga, Fujii:2015jha, Kanemura:2017wtm}. 

This manual of {\tt H-COUP\_1.0} is organized as follows. 
In Sec.~\ref{sec:models}, we define the Higgs potential in the HSM, the THDMs and the IDM.
Constraints implemented in {\tt H-COUP\_1.0} are discussed in Sec.~\ref{sec:constraints}. 
In Sec.~\ref{sec:reno_vertex}, we give the renormalized form factors of the Higgs boson vertices. 
In Sec.~\ref{sec:H-COUP}, the structure of {\tt H-COUP\_1.0} is shown, where the input and output parameters are explained for each model. 
In Sec.~\ref{sec:how_to_run}, we explain how to install and run {\tt H-COUP\_1.0} in order. A sample of output values is presented. 
Sec.~\ref{sec:decay_rate} is devoted to the discussion of the application of {\tt H-COUP\_1.0} to the decay rates.  
We also briefly review QCD corrections to decay rates of $h \to q\bar{q}$ and $h \to gg$ processes.     
Summary is given in Sec.~\ref{sec:summary}.

\section{Non-minimal Higgs sectors}\label{sec:models}

We briefly introduce non-minimal Higgs sectors implemented in {\tt H-COUP\_1.0}, 
i.e. the HSM, the THDMs with a softly-broken $Z_2$ symmetry and CP-conservation and the IDM. 
Throughout this manual, we use the following notation to represent mass eigenstates of fields: 
\begin{align}
\begin{array}{ll}
G^\pm~(G^0)&: \text{the Nambu-Goldstone bosons absorbed by the $W_L^\pm$ and $(Z_L)$ boson}  \\
H,~h&: \text{CP-even Higgs bosons. The latter corresponds to the discovered one} \\
A &:\text{a CP-odd Higgs boson} \\
H^\pm &:\text{a pair of singly-charged Higgs bosons} 
\end{array}
\end{align}
In addition, we use the shorthand notations $s_\theta=\sin\theta$ and $c_\theta=\cos\theta$. 
\subsection{HSM}
\label{HSM}

The Higgs sector of the HSM is composed of an isospin doublet Higgs field $\Phi$ with hypercharge $Y=1/2$ and a real singlet Higgs field $S$ with $Y=0$. 
The most general Higgs potential is written as
 \begin{align}
V(\Phi,S) =&\, m_\Phi^2|\Phi|^2+\lambda |\Phi|^4  
+\mu_{\Phi S}^{}|\Phi|^2 S+ \lambda_{\Phi S} |\Phi|^2 S^2 
+t_S^{}S +m^2_SS^2+ \mu_SS^3+ \lambda_SS^4,\label{Eq:HSM_pot}
\end{align}
where all the parameters are real. Component fields are expressed by 
\begin{align}
\Phi=\left(\begin{array}{c}
G^+\\
\frac{v+\phi+iG^0}{\sqrt{2}}
\end{array}\right),\quad
S=v_S^{} + s. \label{hsm_f}
\end{align}
The Vacuum Expectation Value (VEV) of the doublet field $v$ is related to the electroweak symmetry breaking, namely, $v=(\sqrt{2}G_F)^{-1/2}\simeq 246~{\rm GeV}$ with 
$G_F$ being the Fermi constant, while the VEV of the singlet field $v_S$ does not break any symmetry. 
The potential given in Eq.~(\ref{Eq:HSM_pot}) is invariant under the shift of the singlet VEV $v_S^{} \to v_S'$~\cite{Chen:2014ask}, so that 
$v_S$ can be fixed to be zero without any loss of generality, and we take it in what follows. 

The mass eigenstates of the Higgs bosons are defined by introducing the mixing angle $\alpha$ as follows
\begin{align}
\begin{pmatrix}
s \\
\phi
\end{pmatrix} = R(\alpha)
\begin{pmatrix}
H \\
h
\end{pmatrix}~~\text{with}~~R(\theta) = 
\begin{pmatrix}
c_\theta & -s_ \theta \\
s_\theta & c_\theta
\end{pmatrix}.   \label{mat_r}
\end{align}
Their squared masses and the mixing angle $\alpha$ are expressed as
\begin{align}
 &m_H^2=M_{11}^2c^2_\alpha +M_{22}^2s^2_\alpha +M_{12}^2s_{2\alpha} , \label{mbh}\\
 &m_h^2=M_{11}^2s^2_\alpha +M_{22}^2c^2_\alpha -M_{12}^2s_{2\alpha} , \label{mh}\\
 &\tan 2\alpha=\frac{2M_{12}^2}{M_{11}^2-M_{22}^2}, \label{tan2a}
 \end{align}  
where the mass matrix elements $M^2_{ij}$ are given by
 \begin{align}
M^2_{11}= 2m_S^2+ v^2\lambda_{\Phi S}  ,\quad M^2_{22}=2\lambda v^2,\quad M^2_{12}=v\mu_{\Phi S}^{}. \label{mij}
\end{align}
We note that the parameters $m_\Phi^2$ and $t_S^{}$ are eliminated by using the stationary conditions for $\phi$ and $s$. 
By using Eqs.~\eqref{mbh} - \eqref{tan2a}, we can replace the parameters $\lambda, m_S^2$ and $\mu_{\Phi S}$ with $m_H^2$, $m_h^2$ and $\alpha$. 
As a result, there are the following 5 input free parameters in the HSM: 
\begin{align}
&m_H,~~\alpha,~~\lambda_S,~~\lambda_{\Phi S},~~\mu_{S}. 
\end{align}
We note that $m_h$ and $v$ are fixed to be 125 GeV and by $(\sqrt{2} G_F)^{-1/2}$, respectively, 
and these values are also used in the THDMs and the IDM discussed in the succeeding subsections.

\subsection{THDMs}

\begin{table}[t]
\begin{center}
{\renewcommand\arraystretch{1.2}
\begin{tabular}{c|ccccccc|ccc}\hline\hline
&$\Phi_1$&$\Phi_2$&$Q_L$&$L_L$&$u_R$&$d_R$&$e_R$&$\zeta_u$ &$\zeta_d$&$\zeta_e$ \\\hline
Type-I &$+$&
$-$&$+$&$+$&
$-$&$-$&$-$&$\cot\beta$&$\cot\beta$&$\cot\beta$ \\\hline
Type-II&$+$&
$-$&$+$&$+$&
$-$
&$+$&$+$& $\cot\beta$&$-\tan\beta$&$-\tan\beta$ \\\hline
Type-X &$+$&
$-$&$+$&$+$&
$-$
&$-$&$+$&$\cot\beta$&$\cot\beta$&$-\tan\beta$ \\\hline
Type-Y &$+$&
$-$&$+$&$+$&
$-$
&$+$&$-$& $\cot\beta$&$-\tan\beta$&$\cot\beta$ \\\hline\hline
\end{tabular}}
\caption{The $Z_2$ charge assignment and the $\zeta_f$ ($f=u,d,e$) factors appearing in Eq.~(\ref{hff}) for each type of Yukawa interactions. 
}
\label{yukawa_tab}
\end{center}
\end{table}

The Higgs sector of the THDMs is composed of two isospin doublet Higgs fields $\Phi_1$ and $\Phi_2$ with $Y=1/2$. 
Under the softly-broken $Z_2$ symmetry, we can define four types of Yukawa interactions~\cite{Barger:1989fj,Grossman:1994jb,Aoki:2009ha} 
as shown in Table~\ref{yukawa_tab} depending on the $Z_2$ charge assignment for the right-handed fermions. 

The Higgs potential is given by
\begin{align}
V(\Phi_1,\Phi_2) &= +m_1^2|\Phi_1|^2+m_2^2|\Phi_2|^2-m_3^2(\Phi_1^\dagger \Phi_2 +\text{h.c.})\notag\\
& +\frac{\lambda_1}{2}|\Phi_1|^4+\frac{\lambda_2}{2}|\Phi_2|^4+\lambda_3|\Phi_1|^2|\Phi_2|^2+\lambda_4|\Phi_1^\dagger\Phi_2|^2
+\frac{\lambda_5}{2}\left[(\Phi_1^\dagger\Phi_2)^2+\text{h.c.}\right]. \label{pot_thdm2}
\end{align}
Although  $m_3^2$ and $\lambda_5$ are generally complex, we take them to be real, by which the Higgs potential is CP-conserving. 
The two doublets $\Phi_1$ and $\Phi_2$ are parameterized as 
\begin{align}
\Phi_i=\left(\begin{array}{c}
w_i^+\\
\frac{v_i+h_i+iz_i}{\sqrt{2}}
\end{array}\right),\hspace{3mm}(i=1,2),
\end{align} 
where $v_1$ and $v_2$ are the VEVs of the Higgs doublet fields with $v=\sqrt{v_1^2+v_2^2}$.

The mass eigenstates of the Higgs fields are defined as follows:
\begin{align}
\left(\begin{array}{c}
w_1^\pm\\
w_2^\pm
\end{array}\right)&=R(\beta)
\left(\begin{array}{c}
G^\pm\\
H^\pm
\end{array}\right),\quad 
\left(\begin{array}{c}
z_1\\
z_2
\end{array}\right)
=R(\beta)\left(\begin{array}{c}
G^0\\
A
\end{array}\right),\quad
\left(\begin{array}{c}
h_1\\
h_2
\end{array}\right)=R(\alpha)
\left(\begin{array}{c}
H\\
h
\end{array}\right), \label{mixing}
\end{align}
where $\tan\beta = v_2/v_1$. 
By solving the two stationary conditions for $h_1$ and $h_2$, we can eliminate the parameters $m_1^2$ and $m_2^2$. 
Then, the squared masses of the physical Higgs bosons and the mixing angle $\alpha$ are expressed by 
\begin{align}
&m_{H^\pm}^2 = M^2-\frac{v^2}{2}(\lambda_4+\lambda_5),\label{mass0} \\
 &m_A^2=M^2-v^2\lambda_5, \label{mass1} \\ 
&m_H^2=c^2_{\beta-\alpha} M_{11}^2 + s^2_{\beta-\alpha} M_{22}^2 - s_{2(\beta-\alpha)} M_{12}^2, \label{111}\\
&m_h^2=s^2_{\beta-\alpha} M_{11}^2 + c^2_{\beta-\alpha} M_{22}^2 + s_{2(\beta-\alpha)}M_{12}^2,  \label{222}\\
&\tan 2(\beta-\alpha)= -\frac{2M_{12}^2}{M_{11}^2-M_{22}^2}, \label{333}
\end{align} 
where $M_{ij}^2$ ($i,j=1,2$) are the mass matrix elements for the CP-even scalar states in the basis of $(h_1,h_2)R(\beta)$:
\begin{align}
M_{11}^2&=v^2(\lambda_1c^4_\beta+\lambda_2s^4_\beta)+\frac{v^2}{2}\lambda_{345}s^2_{2\beta},  \label{m11}  \\
M_{22}^2&=M^2+v^2s^2_\beta c^2_\beta(\lambda_1+\lambda_2-2\lambda_{345}), \label{m22}  \\
M_{12}^2&=\frac{v^2}{2} s_{2\beta}( \lambda_2s^2_\beta  -\lambda_1c^2_\beta)+\frac{v^2}{2}s_{2\beta} c_{2\beta} \lambda_{345},  \label{m12}
\end{align}
with $\lambda_{345}\equiv \lambda_3+\lambda_4+\lambda_5$. 
The parameter $M^2 \equiv m_3^2/(s_\beta c_\beta)$ describes the soft breaking scale of the $Z_2$ symmetry. 
From the above discussion, we can choose the following 7 parameters as input free parameters
\begin{align}
m_H^{},~ m_A^{},~ m_{H^\pm},~ s_{\beta-\alpha},~ \tan\beta,~ M^2,~ {\rm Sign}(c_{\beta-\alpha}), 
\end{align} 
where $s_{\beta-\alpha}\geq 0 $ is taken by definition. 

\subsection{IDM}

Similar to the THDMs, the Higgs sector of the IDM is composed of two doublet scalar fields $\Phi$ and $\eta$, but the $Z_2$ symmetry is assumed to be unbroken. 
Under the $Z_2$ symmetry, $\eta$ is $Z_2$ odd, while all the other fields are $Z_2$ even. 
In order to avoid the spontaneous breaking of the $Z_2$ symmetry, the VEV of $\eta$ must be taken to zero, and thus $\eta$ is called the inert doublet. 
The general Higgs potential under the exact $Z_2$ symmetry is written by
\begin{align}
V(\Phi,\eta)&=\mu_1^2|\Phi|^2+\mu_2^2|\eta|^2+\frac{\lambda_1}{2}|\Phi|^4+\frac{\lambda_2}{2}|\eta|^4
+\lambda_{3}|\Phi|^2|\eta|^2+\lambda_{4}|\Phi^\dagger\eta|^2+\frac{\lambda_5}{2}[(\Phi^\dagger\eta)^2+{\rm h.c.}]. \label{IDM_po}
\end{align}
All the parameters are taken to be real without loss of generality.  
Component fields of $\Phi$ and $\eta$ are expressed by
\begin{align}
\Phi=
\begin{pmatrix}
G^+ \\
\frac{v+ h+iG^0}{\sqrt{2}}
\end{pmatrix},\ \ 
\eta =
\begin{pmatrix}
H^+ \\
\frac{H+iA}{\sqrt{2}}
\end{pmatrix},
\end{align}

Imposing a stationary condition for $h$, we can eliminate $\mu^2_1$. 
Then one obtains the following expressions for the squared masses of the Higgs bosons
\begin{align}
m_h^2&=\lambda_{1}v^2, \\
m_{H^\pm}^2&=\mu_2^2+\frac{v^2}{2}\lambda_{3}, \\
m_H^2&=\mu_2^2+\frac{v^2}{2}(\lambda_{3}+\lambda_{4}+\lambda_{5}), \\ 
m_A^2&=\mu_2^2+\frac{v^2}{2}(\lambda_{3}+\lambda_{4}-\lambda_{5}). 
\end{align}
We can choose the following 5 parameters as input free parameters:
\begin{align}
m_H,~ m_A,~ m_{H^\pm},~ \mu_2,~ \lambda_2.
\end{align}

\section{Constraints}\label{sec:constraints}

In this section, we discuss the constraints on the parameter space which are implemented in {\tt H-COUP\_1.0}. 
\begin{itemize}
 \item {\bf Perturbative unitarity bounds}
 
Perturbative unitarity constraints \cite{Lee:1977eg} can be expressed by the following inequality 
  \begin{align}
 |a_0^i|<\frac{1}{2}, 
 \label{unitary_condition}
 \end{align}
where $a_0^i$ are the eigenvalues of the $s$-wave amplitude matrix for the 
2-body $\to$ 2-body elastic scalar boson scatterings at the high-energy limit. 
These eigenvalues are expressed by linear combinations of quartic couplings of the Higgs potential, and it turns out
to be the constraint on the masses of Higgs bosons. 
Originally, perturbative unitarity constraints were applied in the SM to obtain the upper limit on the Higgs boson mass~\cite{Lee:1977eg}, 
where there are 4 neutral 2-body scattering states: $|G^+G^-\rangle,~ |G^0G^0\rangle,~ |G^0h\rangle$ and $|hh\rangle$. 
The eigenvalues $a_0^i$ in Eq.~(\ref{unitary_condition}) can be obtained by diagonalizing the $4\times 4$ $s$-wave amplitude matrix. 
In the HSM, the $s$-wave amplitude matrix for the neutral 2-body scattering states 
is the $7\times 7$ form, which are constructed by $|G^+G^-\rangle$, $|G^0G^0\rangle$, $|G^0h\rangle$, $|G^0H\rangle$, $|hH\rangle$, $|hh\rangle$ and $|HH\rangle$. 
Diagonalizing the $s$-wave amplitude matrix, we can obtain the 4 independent eigenvalues \cite{Cynolter:2004cq}. 
In the THDMs and the IDM, the $s$-wave amplitude matrix for the neutral 2-body scattering states is given by the $14\times 14$ form by 
$|G^+G^-\rangle$, $|G^0G^0\rangle$, $|G^0h\rangle$, $|G^0H\rangle$, $|hH\rangle$, $|hh\rangle$, $|HH\rangle$, 
$|AA\rangle$, $|AG^0\rangle$, $|Ah\rangle$, $|AH\rangle$, $|H^+G^-\rangle$, $|G^-H^+\rangle$ and $|H^+H^-\rangle$.  
This provides the 12 independent eigenvalues~\cite{Kanemura:1993hm,Akeroyd:2000wc,Ginzburg:2005dt,Kanemura:2015ska}.

\item {\bf Triviality bounds}
 
Triviality bounds require that a Landau pole does not appear up to a cutoff scale $\Lambda_{\rm cutoff}$. 
This requirement can be expressed as 
  \begin{align}
 |\lambda_{i}(\mu)|\leq 4\pi, ~~~\text{for}~~~\mu^\forall~~ (m_Z\leq\mu\leq\Lambda_{\rm cutoff} ), 
 \end{align}
where $\lambda_i(\mu)$ are running coupling constants at scale $\mu$ evaluated by solving renormalization group equations. 
The initial scale is fixed to be $\mu = m_Z^{}$. 
In {\tt H-COUP\_1.0}, running couplings are calculated by using one-loop $\beta$ functions 
given in Ref.~\cite{Gonderinger:2009jp}, Ref.~\cite{Inoue:1982ej} and Ref.~\cite{Goudelis:2013uca} for the HSM, the THDM and the IDM, respectively.
 
 \item {\bf Vacuum stability bounds}
  
Vacuum stability bounds require that the Higgs potential is bound from below in any direction of field space with a large value.
The explicit formulae for the necessary and sufficient conditions for the vacuum stability are 
given in Refs.~\cite{Pruna:2013bma, Fuyuto:2014yia, Robens:2015gla} for the HSM 
and in Refs.~\cite{Deshpande:1977rw,Klimenko:1984qx,Sher:1988mj,Kanemura:1999xf} for the THDMs and the IDM. 
In {\tt H-COUP\_1.0}, such condition, written in terms of the running dimensionless coupling constants,  
is required to be satisfied at arbitrary scale $\mu$ with  $m_Z\leq\mu\leq\Lambda_{\rm cutoff}$. 

 \item {\bf Conditions to avoid wrong vacuum (only for the HSM)}

In the HSM, in addition to the true local vacuum, 
$(\sqrt{2}\langle\Phi\rangle,\langle S\rangle)$=$(v,0)$, 
there are 5 wrong local extrema at $(v_\pm, x_\pm)$ and ($0, x_{1,2,3}$) because of the existence of the scalar trilinear couplings $\mu_{\Phi S}$ and $\mu_S$.~\cite{Chen:2014ask,Espinosa:2011ax}. 
Explicit formulae for $v_\pm, x_\pm$ and $ x_{1,2,3}$ are given in Ref.~\cite{Kanemura:2016lkz}.
The conditions to avoid wrong vacua are then expressed as follows
 \begin{align}
 V_{\rm nor}(v_\pm, x_\pm)>0,\quad  V_{\rm nor}(0, x_{1,2,3})>0,
 \end{align} 
where $V_{\rm nor}$ is the normalized  Higgs potential defined by $V_{\rm nor}(v,0)=0$. 

 \item {\bf Constraints from the $S$ and $T$ parameters } 
 
In {\tt H-COUP\_1.0}, the electroweak $S$ and $T$ parameters proposed by Peskin and Takeuchi \cite{Peskin:1990zt} are calculated in each non-minimal Higgs sector. 
Under $U=0$, $S$ and $T$ are given by the global fit of various electroweak observables in Ref.~\cite{Baak:2012kk} as 
\begin{align}
S=0.05 \pm 0.09,\quad T=0.08 \pm 0.07,
\end{align}
with the correlation coefficient $\rho_{ST}^{} =0.89$. 
Defining new physics contributions to the $S$ and $T$ parameters as
\begin{align}
\Delta S=S_{\rm NP}-S_{\rm SM},\quad   \Delta T=T_{\rm NP}-T_{\rm SM},
\end{align} 
we require $\Delta S$ and $\Delta T$ to be within 95\% CL, i.e., $\chi^2(\Delta S, \Delta T)\leq 5.99$.
We note that we take $m_h=125$ GeV and $m_t=173.21$ GeV as a reference value for the SM prediction $S_{\rm SM}$ and $T_{\rm SM}$.  
The analytic formulae for the new physics contributions to $\Delta S$ and $\Delta T$ are given in Ref.~\cite{Lopez-Val:2014jva} and 
\cite{Toussaint:1978zm,Bertolini:1985ia,Peskin:2001rw,Grimus:2008nb,Kanemura:2011sj} for the HSM and the THDM, respectively. 
Those in the IDM are simply obtained by taking $s_{\beta-\alpha}=1$ in the THDMs.

\end{itemize}

\section{Renormalized vertices}\label{sec:reno_vertex}

\begin{figure}[!t]
\begin{center}
\includegraphics[width=150mm]{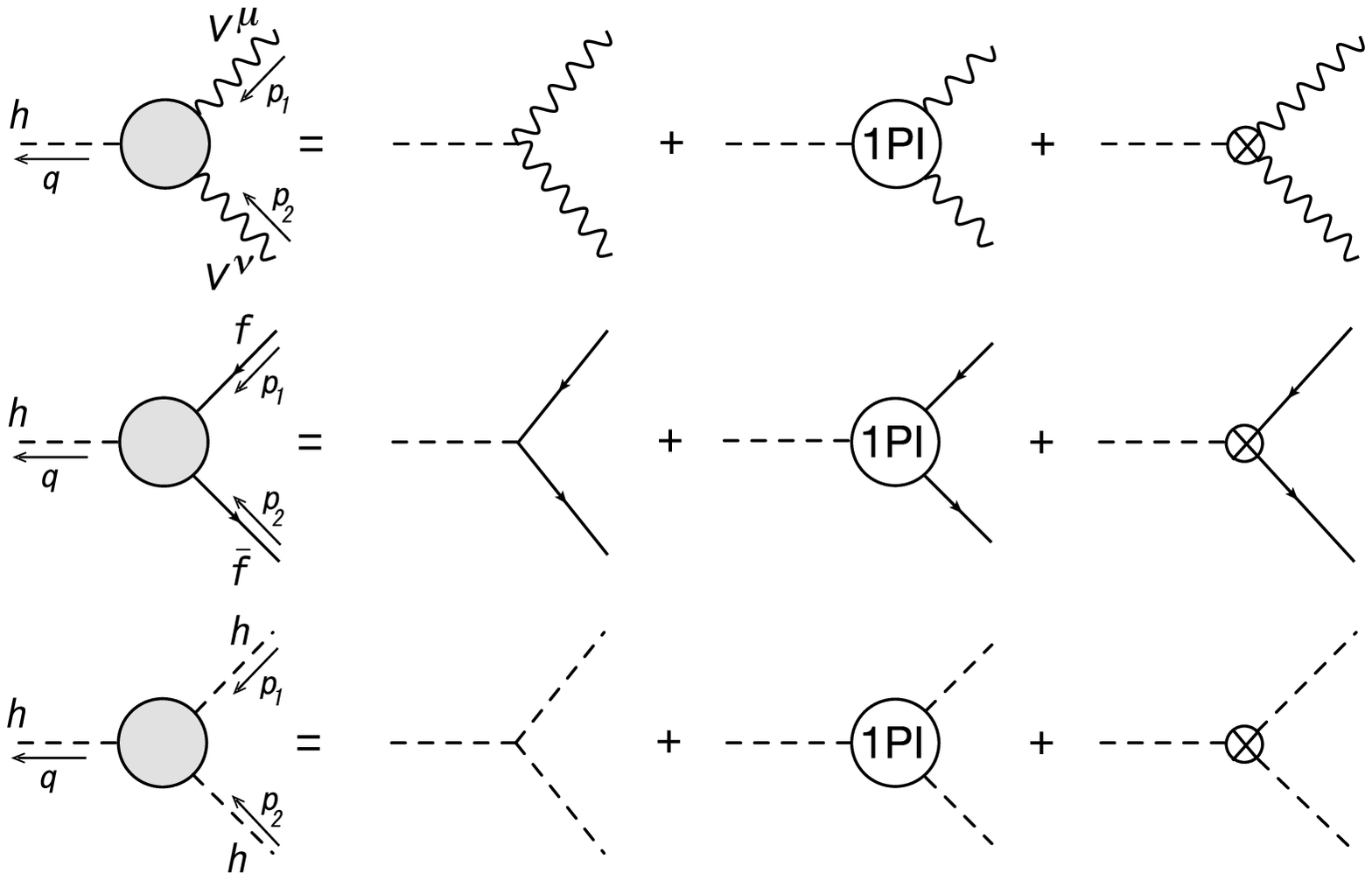}
\caption{Schematic expressions of Eqs. \eqref{eq:renohVV}-\eqref{eq:renohhh}. 
The first, second and third expressions denote the renormalized form factors of the $hVV$ vertex, $hf\bar{f}$ vertex and the $hhh$ vertex function, respectively.
 In  the right hand side of these functions, the first term, second term and third term show contributions at the tree level, at the one-loop level and of the counterterms, respectively.}
\label{Fig:hXX}
\end{center}
\end{figure}

In this section, we define renormalized Higgs boson vertices $hVV$ ($V=W$ or $Z$), $hf\bar{f}$ and $hhh$  at one-loop level, 
which are outputs in {\tt H-COUP\_1.0}.  
We apply the improved on-shell renormalization scheme adopted in Ref.~\cite{Kanemura:2017wtm}, where
gauge dependence appearing in the renormalization of mixing angles among scalar bosons is removed by using the pinch technique. 

The renormalized $hVV$ and $hf\bar{f}$ vertices can be decomposed by the following form factors: 
\begin{align}
\hat{\Gamma}_{h VV}^{\mu\nu}(p_1^2,p_2^2,q^2)&=g^{\mu\nu}\hat{\Gamma}_{h VV}^1
+\frac{p_1^\mu p_2^\nu}{m_V^2}\hat{\Gamma}_{h VV}^2
+i\epsilon^{\mu\nu\rho\sigma}\frac{p_{1\rho} p_{2\sigma}}{m_V^2}\hat{\Gamma}_{h VV}^3,  \label{form_factor} \\
\hat{\Gamma}_{h f\bar{f}}(p_1^2,p_2^2,q^2)&=
\hat{\Gamma}_{h f\bar{f}}^S+\gamma_5 \hat{\Gamma}_{h f\bar{f}}^P+p_1\hspace{-3.5mm}/\hspace{2mm}\hat{\Gamma}_{h f\bar{f}}^{V_1}
+p_2\hspace{-3.5mm}/\hspace{2mm}\hat{\Gamma}_{h f\bar{f}}^{V_2}\notag\\
&\quad +p_1\hspace{-3.5mm}/\hspace{2mm}\gamma_5 \hat{\Gamma}_{h f\bar{f}}^{A_1}
+p_2\hspace{-3.5mm}/\hspace{2mm}\gamma_5\hat{\Gamma}_{h f\bar{f}}^{A_2}
+p_1\hspace{-3.5mm}/\hspace{2mm}p_2\hspace{-3.5mm}/\hspace{2mm}\hat{\Gamma}_{h f\bar{f}}^{T}
+p_1\hspace{-3.5mm}/\hspace{2mm}p_2\hspace{-3.5mm}/\hspace{2mm}\gamma_5\hat{\Gamma}_{h f\bar{f}}^{PT}, 
\end{align}
where the arguments $(p_1^2,p_2^2,q^2)$ for each form factor are understood. 
The direction of the momenta $p_1^\mu$, $p_2^\mu$ and $q^\mu$ for each vertex is shown in Fig.~\ref{Fig:hXX}. 
These renormalized form factors $\hat{\Gamma}^i_{hVV}$ and $\hat{\Gamma}^a_{hf\bar{f}}$ 
and the renormalized $hhh$ vertex are further expressed by the three parts:
\begin{align}
\hat{\Gamma}^i_{hVV}(p_1^2,p_2^2,q^2)&=\Gamma^{i,{\rm tree}}_{hVV}+\Gamma^{i,{\rm 1PI}}_{hVV}(p_1^2,p_2^2,q^2)+\delta \Gamma^{i}_{hVV},\label{eq:renohVV} \\ 
\hat{\Gamma}^a_{h f\bar{f}}(p_1^2,p_2^2,q^2)&=\Gamma^{a,{\rm tree}}_{h f\bar{f}}+\Gamma^{a,{\rm 1PI}}_{h f\bar{f}}(p_1^2,p_2^2,q^2)+\delta \Gamma^{a}_{h f\bar{f}}\ \ ,\label{eq:renohff} \\
\hat{\Gamma}_{hhh}(p_1^2,p_2^2,q^2)&=\Gamma^{{\rm tree}}_{hhh}+\Gamma^{{\rm 1PI}}_{hhh}(p_1^2,p_2^2,q^2)+\delta \Gamma^{}_{hhh},\label{eq:renohhh} 
\end{align}
where $\Gamma_{hXX}^{\rm tree}$, $\Gamma_{hXX}^{\rm 1PI}$ and $\delta\Gamma_{hXX}$ denote 
the contributions from the tree level diagram, 1PI diagrams for the vertex and the counterterms, respectively. 
This can be schematically expressed as in Fig.~\ref{Fig:hXX}. 
The tree level contributions are expressed as 
 \begin{align}
  \Gamma_{hVV,{\rm HSM}}^{1,{\rm tree}}& = \frac{2m_V^2}{v}c_\alpha, \\ 
  \Gamma_{hf\bar{f},{\rm  HSM}}^{S,{\rm tree}}&=-\frac{m_f}{v}c_\alpha, \\ 
  \Gamma_{hhh,{\rm HSM}}^{\rm tree}&=6\left[-\frac{c_\alpha^3}{2v}m_h^2-s_\alpha^2(c_\alpha\lambda_{\Phi S}v-s_\alpha\mu_S)\right],
 \end{align}
in the HSM, 
 \begin{align}
  \Gamma_{hVV,{\rm THDM}}^{1,{\rm tree}}& = \frac{2m_V^2}{v}s_{\beta-\alpha}, \\ 
  \Gamma_{hf\bar{f},{\rm  THDM}}^{S,{\rm tree}}&=-\frac{m_f}{v}(s_{\beta-\alpha}+\zeta_f c_{\beta-\alpha}), \label{hff} \\ 
  \Gamma_{hhh,{\rm THDM}}^{\rm tree}&=6\left[-\frac{m_h^2}{2v}s_{\beta-\alpha}+\frac{M^2-m_h^2}{v}s_{\beta-\alpha}c_{\beta-\alpha}^2+\frac{M^2-m_h^2}{2v}c_{\beta-\alpha}^3(\cot\beta-\tan\beta)\right],
 \end{align} 
in the THDMs, where $\zeta_f$ are defined in Table~\ref{yukawa_tab}.
For the IDM,
 \begin{align}
  \Gamma_{hVV,{\rm IDM}}^{1,{\rm tree}}& = \frac{2m_V^2}{v}, \ \ 
  \Gamma_{hf\bar{f},{\rm  IDM}}^{S,{\rm tree}}=-\frac{m_f}{v},  \ \ 
  \Gamma_{hhh,{\rm IDM}}^{\rm tree}=-3\frac{m_h^2}{v},
 \end{align}  
 where these expressions are the same as those in the SM. We note that tree level contributions to all the other form factors are zero, namely, 
 \begin{align}
 \Gamma_{hVV}^{2,{\rm tree}}=\Gamma_{hVV}^{3,{\rm tree}}=\Gamma_{hf\bar{f}}^{a,{\rm tree}}=0 \ \ (a\neq S).
 \end{align}
Explicit formulae for $\Gamma_{hXX}^{\rm 1PI}$ in the HSM, THDMs and IDM are presented in Refs.~\cite{, Kanemura:2016lkz,Kanemura:2015fra}, \cite{Kanemura:2015mxa} and \cite{Kanemura:2016sos}, respectively, 
and those for counterterms in each model are given in Ref.~\cite{Kanemura:2017wtm}. 

In {\tt H-COUP\_1.0}, one also can obtain leading order (LO) values of the loop induced decay rates 
$h \to \gamma\gamma$, $h \to Z\gamma$ and $h \to gg$. Their explicit analytic formulae are given in Refs.~\cite{Kanemura:2015mxa} (THDMs), 
\cite{Kanemura:2015fra} (HSM) and \cite{Kanemura:2016sos} (IDM).

\section{Structure of H-COUP}\label{sec:H-COUP}

\begin{figure}[!t]
\begin{center}
\includegraphics[width=150mm]{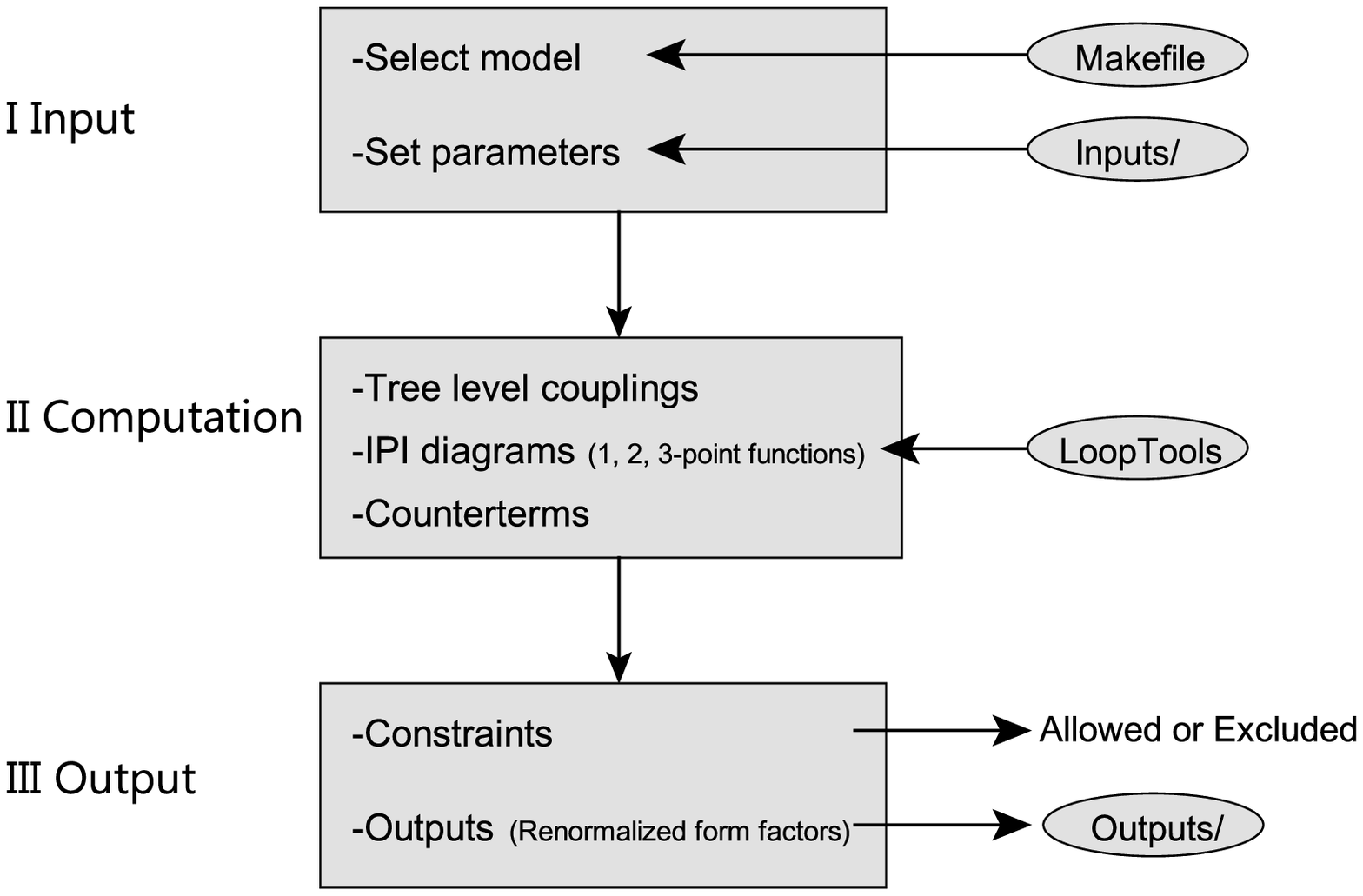}
\caption{Structure of {\tt H-COUP\_1.0. }}
\label{h2}
\end{center}
\end{figure}

\begin{table}[t]
\begin{tabular}{|c|| c c c c c c|}\hline
                  & \multicolumn{6}{c|}{HSM}     \\\hline\hline
Parameters        & $m_H^{}$ & $\alpha$  & $\mu_S^{}$ & $\lambda_S$ & $\lambda_{\Phi S}^{}$ & $\Lambda$ \\\hline 
{\tt H-COUP} def. & {\sf mbh} & {\sf alpha} & {\sf mu\_s}  & {\sf lam\_s} & {\sf lam\_phis} & {\sf cutoff} \\\hline
Default value     & 500 GeV     & 0.1    & 0    & 0     & 0     & 3 TeV     \\\hline
\end{tabular}
\caption{Input parameters in the HSM.  All these parameters are defined by double precision.  }
\label{input_hsm}\vspace{5mm}
\begin{tabular}{|c|| c c c c c c c c c|}\hline
                  & \multicolumn{9}{c|}{THDM}     \\\hline\hline
Parameters        & Type & $m_{H^\pm}^{}$ & $m_A^{}$ & $m_H^{}$ & $M^2$ & $s_{\beta-\alpha}$ & $\text{Sign}(c_{\beta-\alpha})$  & $\tan\beta$ & $\Lambda$\\\hline 
{\tt H-COUP} def. &  {\sf n}   & {\sf mch}          & {\sf ma}      & {\sf mbh}      & {\sf bmsq}  & {\sf sin\_ba}                  & {\sf sign} (1 or $-1$)   & {\sf tanb} & {\sf cutoff}\\\hline 
Default value     &  1   & 500 GeV     & 500 GeV      & 500 GeV       & (450 GeV)$^2$& 0 & 1 & 1.5 & 3 TeV        \\\hline
\end{tabular}
\caption{Input parameters in the THDMs. 
All these parameters are defined by double precision except for 
{\sf n} and {\sf sign} which are defined by integer, and can be ($1,2,3$ or 4) for {\sf n} (see Eq.~(\ref{type})) and (1 or $-1$) for {\sf sign}. }
\label{input_thdm}\vspace{5mm}
\begin{tabular}{|c||c c c c cc|}\hline
                  & \multicolumn{6}{c|}{IDM}     \\\hline\hline
Parameters        &  $m_{H^\pm}^{}$ & $m_A^{}$ & $m_H^{}$ & $\mu_2^2$ & $\lambda_2$ & $\Lambda$ \\\hline 
{\tt H-COUP} def. &  {\sf mch}    & {\sf ma}      & {\sf mbh}      &  {\sf mu2sq}    & {\sf lam2}        & {\sf cutoff}  \\\hline 
Default value     &  500 GeV      & 500 GeV       & 500 GeV       &  (500 GeV)$^2$   & 0                 & 3 TeV \\\hline
\end{tabular}
\caption{Input parameters in the IDM. All these parameters are defined by double precision.  }
\label{input_idm}
\end{table}

\begin{table}[t]
\begin{tabular}{|c||ccc|ccc|}\hline
 &  \multicolumn{3}{c|}{$\hat{\Gamma}^i_{hVV}$ ($V=Z,W$)}      & \multicolumn{3}{c|}{$\hat{\Gamma}^a_{hff}$  ($f=t,b,c,\tau$)}    \\\hline\hline
Momentum     &  $\sqrt{p_1^2}$ & $\sqrt{p_2^2}$ & $\sqrt{q^2}$ &  $\sqrt{p_1^2}$ & $\sqrt{p_2^2}$ & $\sqrt{q^2}$                    \\\hline
{\tt H-COUP} def.   &  {\sf p\_hVV(1) }         & {\sf p\_hVV(2) }  & {\sf p\_hVV(3) }   &  {\sf p\_hff(1) }         & {\sf p\_hff(2) }  & {\sf p\_hff(3) }   \\\hline
Default value       &  $m_V^{}$      & 250 GeV  & $m_h$       &  $m_f$ & $m_f$ (300 GeV) & $m_h$                                                      \\\hline\hline
 &  \multicolumn{3}{c|}{$\hat{\Gamma}_{hhh}$}          \\\cline{1-4}\cline{1-4}
Momentum    &  $\sqrt{p_1^2}$ & $\sqrt{p_2^2}$ & $\sqrt{q^2}$  \\\cline{1-4}
{\tt H-COUP} def.   &  {\sf p\_hhh(1) }         & {\sf p\_hhh(2) }  & {\sf p\_hhh(3) }    \\\cline{1-4}
Default value       &   $m_h$    & $m_h$ & $2m_h$  \\\cline{1-4}
\end{tabular}
\caption{Input global parameters. For $\hat{\Gamma}_{htt}^a$, the default value of $\sqrt{p_2^2}$ ({\sf p\_htt(2) }) is taken to be 300 GeV.  }
\label{global}
\end{table}

\begin{table}[t]
{\footnotesize
\begin{tabular}{|c||ccccc|}\hline
Parameters          &  $m_Z$       & $\alpha_{\text{em}}$     & $G_F$                                & $\Delta \alpha_{\text{em}}$     &  $\alpha_s$       \\\hline 
{\tt H-COUP} def.   &  {\sf mz}    & {\sf alpha\_em}        & {\sf G\_F}                           & {\sf del\_alpha}              & {\sf alpha\_s}          \\\hline 
Description         &  $Z$ mass    & Fine structure const.  & Fermi const.                         &  Shift of $\alpha_{\text{em}}$  & Strong coupling     \\\hline 
Default value       &  91.1876 GeV & 1/137.035999074        & 1.1663787$\times 10^{-5}$ GeV$^{-2}$   & 0.06635                      & 0.1185              \\\hline\hline 
Parameters        & $m_h$                  & $m_t$       & $m_b$           & $m_c$      & $m_\tau$    \\\hline 
{\tt H-COUP} def. & {\sf mh}                    & {\sf mt}    & {\sf mb}        & {\sf mc}  & {\sf mtau}  \\\hline 
Description       & Higgs mass         & $t$ mass    & $b$ mass        & $c$ mass  & $\tau$ mass  \\\hline 
Default value     & 125 GeV                      & 173.21 GeV  &  4.66 GeV       & 1.275 GeV & 1.77684 GeV  \\\hline 
\end{tabular}
}
\caption{Input global SM parameters. All these parameters are defined by double precision.}
\label{sm}
\end{table}

{\tt H-COUP} is composed of the three blocks, i.e., (I) input, (II) computation and (III) output as shown in Fig.~\ref{h2}. 
In the following, we explain each of these blocks in order. 

First, in the input block, we can select the model by modifying Makefile (see the next section). 
In the current version ({\tt H-COUP\_1.0}), we can select the HSM, the THDMs (Type-I, Type-II, Type-X and Type-Y) or the IDM.  
Next, we can set the input parameters which are separated into the model dependent parameters and the global parameters. 
For the former, there 6, 9 and 6 parameters in the HSM, THDMs and IDM as shown in Tables~\ref{input_hsm}, \ref{input_thdm} and \ref{input_idm}, respectively. 
Particularly in the THDM, the type of Yukawa interactions can be specified by setting the {\sf n} variable defined in {\tt H-COUP} as follows:
\begin{align}
n=1:~ \text{Type-I},\quad 
n=2:~ \text{Type-II},\quad 
n=3:~ \text{Type-X}, \quad
n=4:~ \text{Type-Y}.  \label{type}
\end{align}
On the other hand, 
the global parameters are common to all the models, 
by which we can specify the SM parameters and the squared momenta of the renormalized form factors. 
There are 3 independent squared momenta for each renormalized vertex ($\hat{\Gamma}_{hVV}^i$, $\hat{\Gamma}_{hff}^a$ and $\hat{\Gamma}_{hhh}$) as shown in Table~\ref{global}. 
The SM parameters and their default values are summarized in Table~\ref{sm}. 
We can set these model dependent and global input parameters by modifying the files in the inputs directly  (see the next section). 

Second, in the computation block, tree level Higgs boson couplings, 1PI diagrams for 1-, 2- and 3-point functions
and counterterms are calculated under the fixed model and input parameters. 
All the 1PI diagrams are written in terms of the Passarino-Veltman $A$, $B$ and $C$ functions, which are numerically evaluated by {\tt LoopTools}~\cite{Hahn}.  
Then, all the counterterms are evaluated in terms of the above calculated 1-, 2- and 3-point functions. 

Finally, in the output block, {\tt H-COUP} tells us if a given configuration determined by input parameters is allowed or excluded. 
If it is allowed, a message ``Allowed'' appears in the command line after executing the executable file (see the next section). 
If it is excluded, a message ``Excluded by XXX'' appears, where ``XXX'' can be perturbative unitarity, vacuum stability, triviality, wrong vacuum conditions and/or ST parameters. 
In the both cases, the output file is generated in the output directly. 
We note that the output file for the SM predictions is also generated at the same time once one of the output file for the non-minimal Higgs sector is generated.

\section{Installation and how to run}\label{sec:how_to_run}

\begin{table}
\begin{tabular}{|c||cccccc|}\hline
Outputs           &  $\hat{\Gamma}_{hVV}^i$ & $\hat{\Gamma}_{hff}^a$  & $\hat{\Gamma}_{hhh}$ &  $\Gamma(h\to \gamma\gamma)$ & $\Gamma(h\to Z\gamma)$& $\Gamma(h\to gg)$ \\\hline 
{\tt H-COUP} def. &  {\sf rGam\_hVV(i)}    & {\sf rGam\_hff(a)}   & {\sf rGam\_hhh}  &  {\sf Gam\_hgamgam}  & {\sf Gam\_hZgam}  & {\sf Gam\_hgg} \\\hline
\end{tabular}
\caption{Contents of output file. The index $i$ runs from 1 to 3, and the index $a$ runs over $S$, $P$, $V1$, $V2$, $A1$, $A2$, $T$ and $PT$. }
\label{outaa}
\end{table}

In order to run {\tt H-COUP}, we need to install a Fortran compiler (GFortran is recommended) and {\tt LoopTools}~\cite{Hahn} in advance. 
One can download the {\tt LoopTools} package from \cite{Hahn}, and see the manual for its installation. 
In the following, we explain how to run {\tt H-COUP} in order. 

 \begin{enumerate}
 \item Unzip the HCOUP-1.0.zip file:
 \begin{center}
\fbox{
 \$  unzip HCOUP-1.0.zip}
\end{center}
Then, the HCOUP-1.0 directly (HCOUP-1.0/) is created. In this directly, one can find 4 directories and Makefile as follows
\begin{center}
\fbox{
\begin{tabular}{l}
\$  ls  \\
\quad  Makefile~~~inputs/~~~models/~~~modules/~~~outputs/ 
\end{tabular}
}
\end{center}
These directories include the following files.\\\\
\quad inputs/ : files for the model dependent/global input parameters are stored: \\ 
\quad\quad\quad\quad\quad\quad  in\_hsm.txt      (input file for the HSM), \\
\quad\quad\quad\quad\quad\quad  in\_thdm.txt     (input file for the THDMs), \\
\quad\quad\quad\quad\quad\quad  in\_idm.txt      (input file for the IDM),  \\
\quad\quad\quad\quad\quad\quad  in\_momentum.txt (global input file for momenta) and   \\
\quad\quad\quad\quad\quad\quad  in\_sm.txt       (global input file for the SM parameters).  \\
\quad outputs/: files for the output are generated in this directly\footnote{Initially, this folder is empty. }: \\
\quad\quad\quad\quad\quad\quad  out\_sm.txt (output file for the SM), \\
\quad\quad\quad\quad\quad\quad  out\_hsm.txt (output file for the HSM), \\
\quad\quad\quad\quad\quad\quad  out\_thdm.txt (output file for the THDMs) and \\
\quad\quad\quad\quad\quad\quad  out\_idm.txt (output file for the IDM).  \\
\quad models/: main Fortran90 files for {\tt H-COUP} are stored: \\
\quad\quad\quad\quad\quad\quad  HCOUP\_HSM.F90 (model file for the HSM),  \\
\quad\quad\quad\quad\quad\quad  HCOUP\_THDM.F90 (model file for the THDMs) and  \\
\quad\quad\quad\quad\quad\quad  HCOUP\_IDM.F90 (model file for the HSM).  \\
\quad\quad\quad\quad\quad Users do not need to touch these files. \\
\quad modules/: module files for {\tt H-COUP} are stored. \\ 
\quad\quad\quad\quad\quad Users do not need to touch these files. 
 \item Open Makefile by an editor and replace ``PATH'' appearing 
 in the two lines ``LIBS  = -L PATH -looptools''  and  ``\$ (FC) -I PATH -c  \$ $<$ \$ (LIBS)''
 in Makefile by the correct path to the library file of {\tt LoopTools} (looptools.a) and the header file (looptools.h), respectively. 
 These files are in the LoopTools-X.XX/build/ directly (X.XX denotes the version) of the {\tt LoopTools} package. 
\item 
In HCOUP-1.0/, perform the ``make'' command:
\begin{center}
\fbox{
\$  make
}
\end{center}
The executable file ``a.out'' should be generated. 
\item 
Execute a.out:
\begin{center}
\fbox{ \$  ./a.out}
\end{center}
Then, an output file is generated in the output directly (output/). 
If a given set of input parameters is excluded by some of constraints, a message appears in the command line. 
We here show the example of the generated output file in output/ in Fig.~\ref{outbb}. 
\item One can change the model by replacing the ``MODEL'' part of ``MAIN   = MODEL.F90'' in Makefile by the other model file (as default, it is specified to be HCOUP\_HSM.F90 ). 
One can choose one of the HCOUP\_HSM.F90 (for the HSM),  HCOUP\_THDM.F90 (for the THDM) and HCOUP\_IDM.F90 (for the IDM) files. 
\item One can change the model dependent input parameters by modifying the in\_hsm.txt, in\_thdm.txt and in\_idm.txt, 
and also the global (model independent) parameters by modifying the in\_momentum.txt and in\_sm.txt files in the input directly. 
In Fig.~\ref{in1} and \ref{in2}, we show the example of the input file for the HSM (in\_hsm.txt) and the global input file  (in\_momentum.txt). 

\begin{figure}[!t]
\begin{center}
\includegraphics[width=120mm]{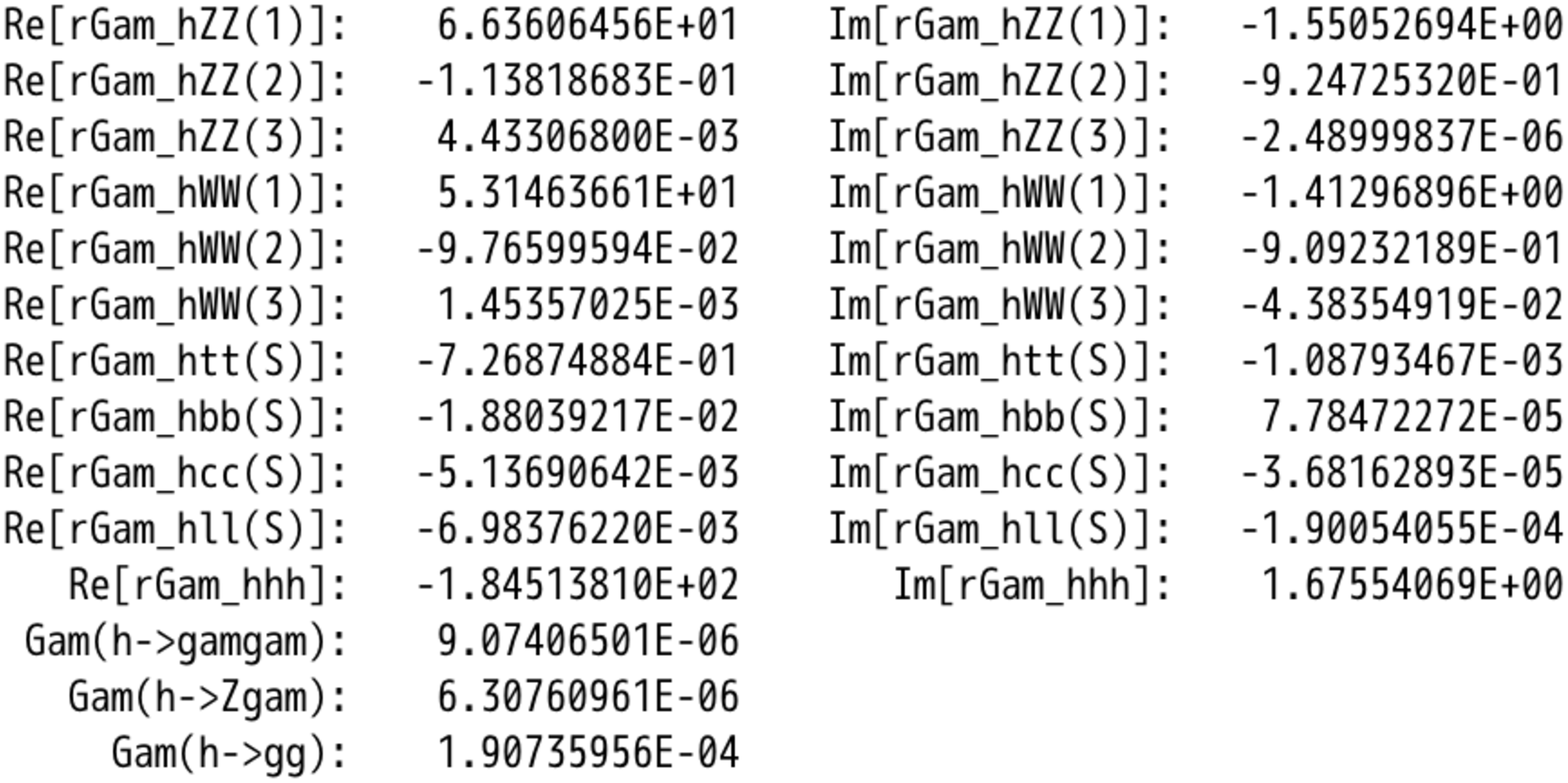}
\caption{Example of the output file (out\_hsm.txt). }. 
\label{outbb}
\end{center}
\end{figure}

\begin{figure}[!t]
\begin{center}
\includegraphics[width=120mm]{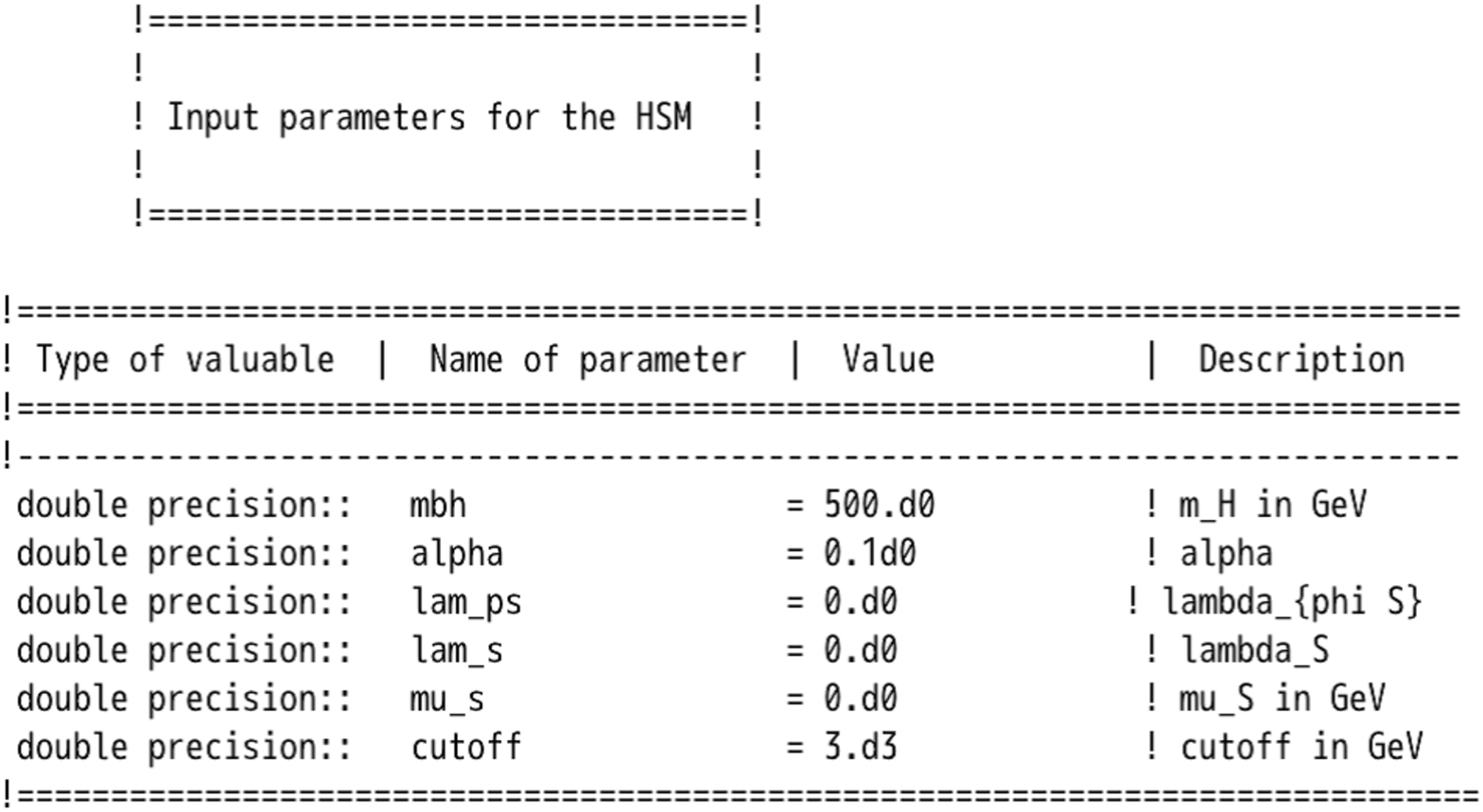}\\
\caption{Example of the input file in\_hsm.txt. } 
\label{in1}
\vspace{5mm}
\includegraphics[width=150mm]{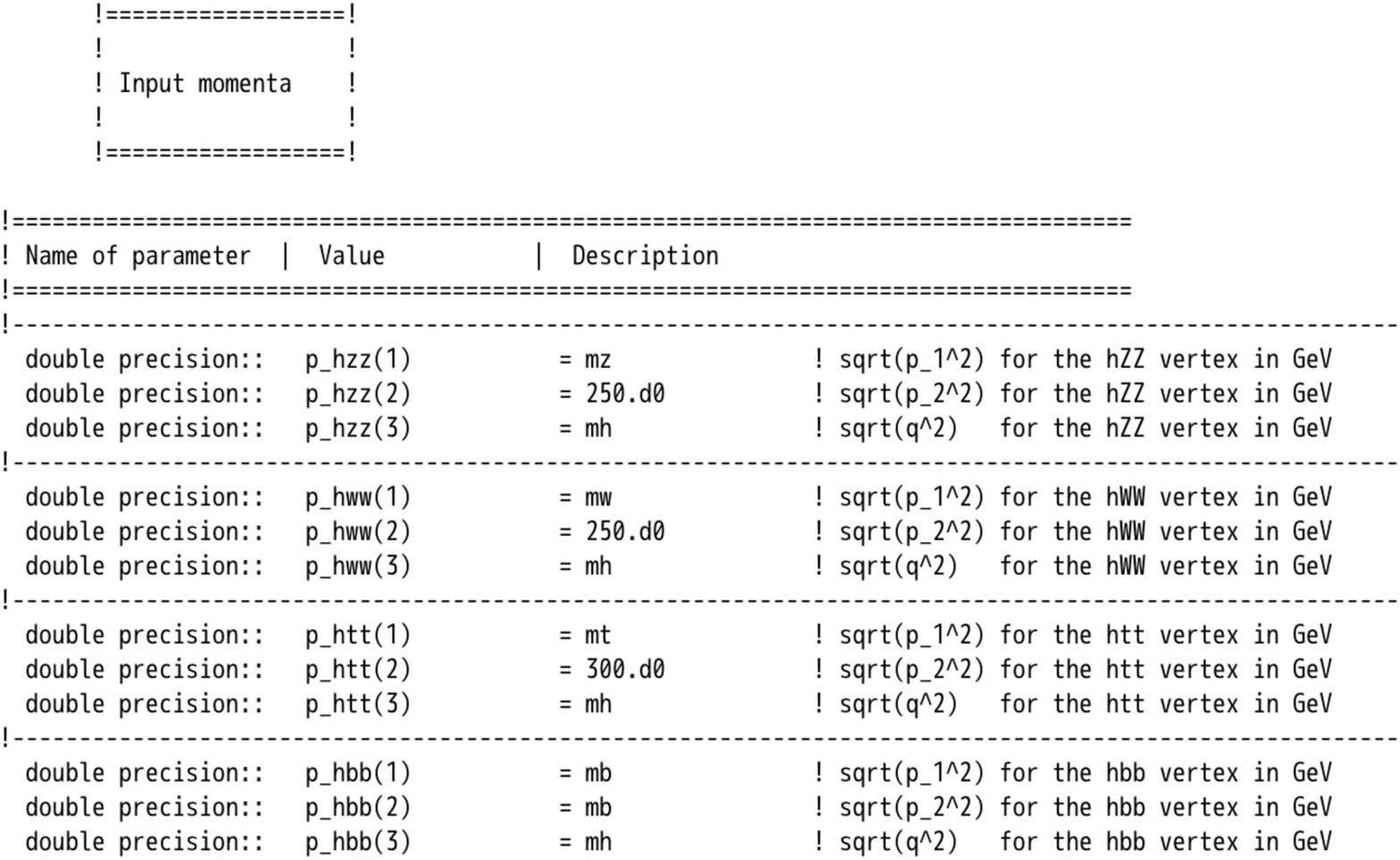}
\caption{Example of the input file in\_momentum.txt. }
\label{in2}
\end{center}
\end{figure}

In Table~\ref{out3}, we show the sample of outputs using the default inputs given in Tables~\ref{input_hsm}, \ref{input_thdm}, \ref{input_idm} and \ref{global}.

\begin{table}
\begin{tabular}{|c||c|c|c|c|}\hline
                               &  SM   &  HSM & THDM (Type-I) & IDM   \\\hline\hline
Re$\hat{\Gamma}_{hZZ}^1$ [GeV]  &  66.7002033       &$66.3606456$       & $66.6615250$     & $66.7002111$         \\\hline
Re$\hat{\Gamma}_{hZZ}^2$ [GeV]  & $-1.14685795\times 10^{-1}$ &$-1.13818683\times 10^{-1}$  & $-1.08763248\times 10^{-1}$     & $-1.14685795\times 10^{-1}$             \\\hline
Re$\hat{\Gamma}_{hZZ}^3$ [GeV]  & $4.45532608\times 10^{-3}$     &$4.43306800\times 10^{-3}$    & $4.45532608\times 10^{-3}$   &$4.45532608\times 10^{-3}$   \\\hline
Re$\hat{\Gamma}_{hWW}^1$ [GeV]  &$53.4180776$       &$53.1463661$      & $53.3946191$    &$53.4180837$\\\hline
Re$\hat{\Gamma}_{hWW}^2$ [GeV]  &$-9.81797093\times 10^{-2}$&$-9.76599594\times 10^{-2}$  & $ -9.26602552\times 10^{-2}$                 &$-9.81797093\times 10^{-2}$\\\hline
Re$\hat{\Gamma}_{hWW}^3$ [GeV]  &$1.46086850\times 10^{-3}$   &$1.45357025\times 10^{-3}$   & $1.46086850\times 10^{-3}$    & $1.46086850\times 10^{-3}$\\\hline
Re$\hat{\Gamma}_{htt}^S$  &  $-7.30596472\times 10^{-1}$  &$-7.26874884\times 10^{-1}$       & $-7.30899549\times 10^{-1}$  & $-7.30596391\times 10^{-1}$\\\hline
Re$\hat{\Gamma}_{hbb}^S$  &  $-1.88986892\times 10^{-2} $ &$-1.88039217\times 10^{-2}$       & $-1.88720530\times 10^{-2}$  &$-1.88986871\times 10^{-2}$\\\hline
Re$\hat{\Gamma}_{hcc}^S$  &   $-5.16279547\times 10^{-3} $   &$-5.13690642\times 10^{-3}$       & $-5.16584879\times 10^{-3}$  &$-5.16279487\times 10^{-3}$\\\hline
Re$\hat{\Gamma}_{h\tau\tau}^S$ & $-7.01896236\times 10^{-3} $&$-6.98376220\times 10^{-3}$   & $-7.02321748\times 10^{-3}$  
&$-7.01896153\times 10^{-3}$\\\hline
Re$\hat{\Gamma}_{hhh}$ [GeV] &  $-1.87380208\times 10^2 $ &$-1.84513810\times 10^{2}$    & $-1.91267505\times 10^{2}$   
&$-1.87380186\times 10^{2}$\\\hline
$\Gamma_{h \to\gamma\gamma}$ [GeV] &  $9.16541404\times 10^{-6}$&$9.07406501\times 10^{-6}$  & $8.95726899\times 10^{-6}$  &$9.16541404\times 10^{-6}$    \\\hline
$\Gamma_{h \to Z\gamma}$ [GeV] &  $6.37110861\times 10^{-6}$&$6.30760961\times 10^{-6}$      & $6.31549430\times 10^{-6}$ &$6.37110861\times 10^{-6}$\\\hline
$\Gamma_{h\to gg}$ [GeV] &  $1.92656103\times 10^{-4}$ &$1.90735956\times 10^{-4}$          & $1.92656103\times 10^{-4}$  &$1.92656103\times 10^{-4}$\\\hline
\end{tabular}
\caption{Sample of the outputs for the renormalized form factors and the loop induced decay rates. 
We note that the QCD corrections are not included in the current version of the {\tt H-COUP} (Ver. 1.0).  
}
\label{out3}
\end{table}

\end{enumerate}


\section{Application of {\tt H-COUP} to physical quantities}\label{sec:decay_rate}

As we explained in the previous sections, {\tt H-COUP} provides numerical values of the renormalized form factors for the $hVV$, $hf\bar{f}$ and $hhh$ vertices. 
Using these form factors, we can compute physical observables, e.g., decay rates, total widths and cross sections. 
We here discuss the application of  {\tt H-COUP} to the calculation of the decay rates of $h$. 
In version 1.0, this cannot be automatically done\footnote{In a future version, physical quantities will also be able to be produced.}, 
but the decay rates can be obtained by the simple computations discussed below. 

\subsection{Decay rates}

Assuming all the extra Higgs bosons are heavier than the discovered Higgs boson $h$, 
there are the following decay processes 
\begin{align}
h \to f\bar{f}~~(f\neq t), \quad h \to VV^* \to Vf\bar{f}', 
\quad h \to \gamma\gamma,\quad h \to Z\gamma ,\quad h \to gg. 
\end{align}
The last three modes are loop induced, and the LO prediction can be provided by {\tt H-COUP\_1.0}.  

The decay rate of $h$ into a fermion pair is given in terms of the form factors of the $hf\bar{f}$ vertex:
\begin{align}
\Gamma(h\to f\bar{f})=\frac{N_c^fm_h}{8\pi}\left|\Gamma_{hff}^S+2m_f\Gamma^{V1}_{hff}+(m_h^2-m_f^2)\Gamma_{hff}^T\right|^2\lambda^{3/2}\left(\frac{m_f^2}{m_h^2},\frac{m_f^2}{m_h^2}\right), 
\end{align}
where $N_c^f = 1 (3)$ for $f$ to be leptons (quarks), and 
\begin{align}
\lambda(x,y) \equiv \left(1+x-y\right)^2-4x. 
\end{align}

In the above expression, the three momenta for the form factors $\Gamma_{hff}^i$ are fixed to be $p_1^2 = p_2^2 = m_f^2$ and $q^2 = m_h^2$. 
We note that the form factors $\Gamma_{hff}^P$, $\Gamma_{hff}^{A1}$, $\Gamma_{hff}^{A2}$ and $\Gamma_{hff}^{PT}$ do not contribute to the decay rate in the case with the on-shell fermions in the final state. 
For the case with $f = q$ ($q$ denotes a light quark such as $b$ and $c$), 
QCD corrections are quite important to be taken into account. 
We will discuss the implementation of QCD corrections into the decay rates of $h \to q\bar{q}$ and $h \to gg$ in the next subsection. 

The decay rate of the $h\to VV^*\to Vf\bar{f}'$ can also be written in terms of the three form factors of the $hVV$ vertex as 
\begin{align}
&\Gamma(h\to VV^*\to Vf\bar{f}')=\frac{g_V^2m_h}{3072\pi^3}\int^{(m_h-m_V)^2}_0\frac{ds}{x_V^3(s-m_V^2)^2}\lambda^{1/2}(x_V,x_s) \notag\\
&\times\Bigg\{ |\Gamma_{hVV}^1|^2x_V^2\left[\left(1-x_V^{}-x_s\right)^2+8x_V^{} x_s\right] \notag\\ 
&+\frac{|\Gamma_{hVV}^2|^2}{4}\left[x_s-(1-\sqrt{x_V^{}})^2\right]\left[x_s-(1+\sqrt{x_V^{}})^2\right]\lambda(x_V,x_s)\notag\\
& +2|\Gamma_{hVV}^3|^2x_V^{}x_s\lambda(x_V,x_s)
+{\rm Re}(\Gamma_{hVV}^{1\ast}\Gamma_{hVV}^2)x_V^{}\left(1-x_V^{}-x_s\right)\lambda(x_V,x_s)  \Bigg\}. \label{hvv}
\end{align}
where $x_V^{} = m_V^2/m_h^2$, $x_s^{} = s/m_h^2$, $g_W^2 = g^2 $ and  $g_Z^2 = g^2(v_f^2 + a_f^2)/c_W^2$ with $v_f=I_f/2-s_W^2Q_f$ and $a_f=I_f/2$. 
We  note that, the masses of the final state fermions are neglected. 
For this process, the squared three momenta in the form factors $\Gamma_{hVV}^i$ should be fixed by $p_1^2 = m_V^2$, $p_2^2 = s$ and $q^2 = m_h^2$, where 
$s$ (the invariant mass of the two fermions) is integrated out from $0$ to $(m_h-m_V)^2$ as seen in Eq.~(\ref{hvv}). 

\subsection{QCD corrections}

It is well known that QCD corrections to the decay rates of $h$ into hadronic final states such as $h \to q\bar{q}$ and $h \to gg$ are quite important. 
Although in {\tt H-COUP\_1.0}, 
QCD corrections are not implemented in the renormalized form factors, it is straightforward to implement such effect in  a future version of {\tt H-COUP} as these
have already been computed in previous works. 

For the  $h \to q\bar{q}$ decays, QCD corrections based on the $\overline{\text{MS}}$ scheme can be expressed at the scale $\mu = m_h$ by 
\begin{align}
\Gamma(h \to q\bar{q}) =\frac{3G_F}{4\sqrt{2}\pi}m_h\bar{m}_q^2(m_h)\left[1 + \sum_{p \geq 1} \Delta \Gamma_p \left(\frac{\bar{\alpha}_s(m_h)}{\pi} \right)^p \right], 
\end{align}
where $\bar{m}_q$ and $\bar{\alpha}_s$ are the running quark mass and the strong coupling constant 
defined in the $\overline{\text{MS}}$ scheme, and the limit $\bar{m}_q(m_h)/m_h \to 0$ is taken. 
The coefficients $\Delta \Gamma_p$ has been known up to $p = 4$ in Ref.~\cite{hqq_alp4}, and their numerical values are given by 
\begin{align}
\Delta\Gamma_1 = 5.6668,\quad
\Delta\Gamma_2 = 29.147,\quad
\Delta\Gamma_3 = 41.758,\quad
\Delta\Gamma_4 = -825.7, 
\end{align}
where the effective quark flavour $n_f = 5$ is taken. 
We note that ${\cal O}(\alpha_s\alpha_{\text{em}})$ corrections to the decay rate of $h \to b\bar{b}$ 
has also been computed in the SM~\cite{hqq_alpmix}, where the magnitude of the correction is comparable to the ${\cal O}(\alpha_s^3)$ correction but has the opposite sign.


For the $h \to gg$ decay, LO is one-loop induced as follows:
\begin{align}
\Gamma(h \to gg)_{\text{LO}} = \frac{G_F\alpha_s^2 m_h^3}{64\sqrt{2}\pi^3 }\left|\sum_q\frac{8m_q^2}{m_h^2}\left[1+\left(2m_q^2 - \frac{m_h^2}{2}\right)C_0(0,0,m_h^2,m_q,m_q,m_q)\right]\right|^2, 
\end{align}
where $C_0$ is the Passarino-Veltman scalar three-point function. 
This can also be applied to obtain the prediction for the Higgs boson production cross section via the gluon fusion mechanism: $gg \to h$ by multiplying the appropriate phase factor. 
The ${\cal O}(\alpha_s)$ correction, corresponding to the next-to-leading order (NLO), to this process has been calculated in Refs.~\cite{hgg_nlo_1,hgg_nlo_2,hgg_nlo_3} in the $\overline{\text{MS}}$
scheme, which gives about $+70\%$ enhancement in the decay rate with respect to the leading order prediction. 
In the limit $m_t \to \infty$, one can obtain a rather simple analytic expression~\cite{djouadi}
\begin{align}
\Gamma(h \to gg)_{\text{NLO}} = \Gamma(h \to gg)_{\text{LO}}\left[1 + \frac{\bar{\alpha}_s(\mu)}{\pi}\left(\frac{95}{4}-\frac{7}{6}n_f + \frac{33-2n_f}{6}\ln \frac{\mu^2}{m_h^2} \right) \right],  
\end{align}
In addition, ${\cal O}(\alpha_s^2)$ corrections  have been given in Refs.~\cite{hgg_nnlo_1,hgg_nnlo_2,hgg_nnlo_3} 
with a limit $m_t \to \infty$ and in Refs.~\cite{hgg_nnlo_mt_1,hgg_nnlo_mt_2} with a finite value of $m_t$. 
In Ref.~\cite{hgg_nnlo_lhc}, the gluon fusion cross section at the LHC with the collision energy of 8 TeV has been presented at ${\cal O}(\alpha_s^2)$ level, and it has been found that 
the theoretical uncertainty is about order $\pm 9\%$. 
Furthermore, ${\cal O}(\alpha_s^3)$ corrections have been performed in Ref.~\cite{hgg_n3lo} (references therein for the other calculations at ${\cal O}(\alpha_s^3)$ level with approximations) 
as the most accurate theoretical prediction in the computation by perturbative QCD.  The theoretical uncertainty in the gluon fusion cross section reduces to be order $\pm 2\%$ at the LHC with 8 TeV. 

These QCD corrections will be implemented in a future version of {\tt H-COUP}.

\section{Summary}\label{sec:summary}

We have described a Fortran program {\tt H-COUP\_1.0} which provides numerical values of various 
one-loop electroweak corrected form factors of the discovered Higgs boson vertices ($hf\bar{f}$, $hVV$ and $hhh$) 
and the loop induced decay rates ($h \to \gamma\gamma$, $h \to  Z\gamma$ and $h \to  gg$) in the SM, HSM, THDMs and IDM. 
The improved on-shell scheme without gauge dependence is applied to perform the renormalization.  
After defining the Higgs potential of each non-minimal Higgs sector, we discussed the independent input parameters and the constraints implemented in {\tt H-COUP\_1.0}, 
i.e., the bounds from the perturbative unitarity, vacuum stability, triviality, wrong vacuum conditions and electroweak $S$ and $T$ parameters. 
The renormalized  $hVV$, $hf\bar{f}$ and $hhh$ vertices are defined in terms of the 3, 8 and 1 form factors, respectively. 
Then, we have explained the structure of {\tt H-COUP\_1.0} and how to install and use it. 
Users can double-check whether {\tt H-COUP\_1.0} correctly works or not by comparing the sample output values presented in Table~\ref{out2}.  

Finally, we are now preparing ... would like to give a list of what we can improve {\tt H-COUP} in its future version as follows:
\begin{enumerate}
\item addition of the calculation for the renormalized vertices for extra Higgs bosons (e.g., $HV^\mu V^\nu$, $Af\bar{f}$), 
\item addition of the other extended Higgs models such as models with isospin triplet Higgs fields, 
\item addition of extra spin 1/2 (e.g., vector-like fermions) and spin 1 (e.g., $Z'$ bosons), 
\item application of our vertex calculations to more physical quantities such as decay rates, branching ratios and cross sections, 
\item inclusion of QCD corrections particularly for the $hq\bar{q}$, $hgg$, $h\gamma\gamma$ and $hZ\gamma$ vertices, 
\item addition of the calculation in the other renormalization schemes such as the $\overline{\text{MS}}$ scheme. 
\end{enumerate}

 
\vspace*{4mm}
\noindent
\section*{Acknowledgments}
\noindent

The authors would like to thank
Fawzi Boudjema, Keisuke Fujii, Howard E. Haber, Mitsuru Kakizaki, Sabine Kraml, Kentarou Mawatari,
Stefano Moretti and Tetsuo Shindou for careful reading of the manual and try to use the beta-version of {\tt H-COUP\_1.0}. 
This work was supported, in part, by Grant-in-Aid for Scientific Research on Innovative Areas, 
the Ministry of Education, Culture, Sports, Science and Technology, No.~16H06492,
Grant H2020-MSCA-RISE-2014 No.~645722 (Non-Minimal Higgs), and JSPS Joint Research Projects (Collaboration, Open Partnership)
``New Frontier of neutrino mass generation mechanisms via Higgs physics at LHC and flavour physics''.



\begin{thebibliography}{1}

\bibitem{LHC_Higgs_ATLAS} 
  G.~Aad {\it et al.} [ATLAS Collaboration],
  Phys.\ Lett.\ B {\bf 716}, 1 (2012)
  [arXiv:1207.7214 [hep-ex]].

\bibitem{LHC_Higgs_CMS} 
  S.~Chatrchyan {\it et al.} [CMS Collaboration],
  Phys.\ Lett.\ B {\bf 716}, 30 (2012)
  [arXiv:1207.7235 [hep-ex]].

\bibitem{Aad:2015zhl} 
  G.~Aad {\it et al.} [ATLAS and CMS Collaborations],
  Phys.\ Rev.\ Lett.\  {\bf 114}, 191803 (2015)
  [arXiv:1503.07589 [hep-ex]].

\bibitem{Khachatryan:2016vau} 
  G.~Aad {\it et al.} [ATLAS and CMS Collaborations],
  JHEP {\bf 1608}, 045 (2016)
  [arXiv:1606.02266 [hep-ex]].

\bibitem{Barklow:2015tja} 
  T.~Barklow, J.~Brau, K.~Fujii, J.~Gao, J.~List, N.~Walker and K.~Yokoya,
  arXiv:1506.07830 [hep-ex].

\bibitem{Tian:2016qlk} 
  J.~Tian {\it et al.} [ILC physics and detector study Collaboration],
  Nucl.\ Part.\ Phys.\ Proc.\  {\bf 273-275}, 826 (2016).

\bibitem{CLIC:2016zwp} 
  M.~J.~Boland {\it et al.} [CLIC and CLICdp Collaborations],
  arXiv:1608.07537 [physics.acc-ph].

\bibitem{Gomez-Ceballos:2013zzn} 
  M.~Bicer {\it et al.} [TLEP Design Study Working Group],
  JHEP {\bf 1401}, 164 (2014)
  [arXiv:1308.6176 [hep-ex]].


\bibitem{hqq_NLO1} 

  E.~Braaten and J.~P.~Leveille,
  Phys.\ Rev.\ D {\bf 22}, 715 (1980).

\bibitem{hqq_NLO2}

  N.~Sakai,
  Phys.\ Rev.\ D {\bf 22}, 2220 (1980).

\bibitem{hqq_NLO3}

  T.~Inami and T.~Kubota,
  Nucl.\ Phys.\ B {\bf 179}, 171 (1981).

\bibitem{hqq_NLO4}

  M.~Drees and K.~i.~Hikasa,
  Phys.\ Rev.\ D {\bf 41}, 1547 (1990).

\bibitem{Fleischer}

  J.~Fleischer and F.~Jegerlehner,
  Phys.\ Rev.\ D {\bf 23}, 2001 (1981).


\bibitem{Kniehl_hff}

  B.~A.~Kniehl,
  Nucl.\ Phys.\ B {\bf 376}, 3 (1992).

\bibitem{Kniehl_hzz}

  B.~A.~Kniehl,
  Nucl.\ Phys.\ B {\bf 352}, 1 (1991).

\bibitem{Kniehl_hww}

  B.~A.~Kniehl,
  Nucl.\ Phys.\ B {\bf 357}, 439 (1991).

\bibitem{hhh-sm1} 
  S.~Kanemura, S.~Kiyoura, Y.~Okada, E.~Senaha and C.~P.~Yuan,
  Phys.\ Lett.\ B {\bf 558}, 157 (2003)
  [hep-ph/0211308].

\bibitem{Kanemura:2004mg} 
  S.~Kanemura, Y.~Okada, E.~Senaha and C.-P.~Yuan,
  Phys.\ Rev.\ D {\bf 70}, 115002 (2004)
  [hep-ph/0408364].


\bibitem{Dabelstein:1995js} 
  A.~Dabelstein,
  Nucl.\ Phys.\ B {\bf 456}, 25 (1995)
  [hep-ph/9503443].

\bibitem{Jimenez:1995wf} 
  R.~A.~Jimenez and J.~Sola,
  Phys.\ Lett.\ B {\bf 389}, 53 (1996)
  [hep-ph/9511292].

\bibitem{Chankowski:1992er} 
  P.~H.~Chankowski, S.~Pokorski and J.~Rosiek,
  Nucl.\ Phys.\ B {\bf 423}, 437 (1994)
  [hep-ph/9303309].

\bibitem{Hollik:2001px} 
  W.~Hollik and S.~Penaranda,
  Eur.\ Phys.\ J.\ C {\bf 23}, 163 (2002)
  [hep-ph/0108245].

\bibitem{Dobado:2002jz} 
  A.~Dobado, M.~J.~Herrero, W.~Hollik and S.~Penaranda,
  Phys.\ Rev.\ D {\bf 66}, 095016 (2002)
  [hep-ph/0208014].

\bibitem{Carena:2001bg} 
  M.~Carena, H.~E.~Haber, H.~E.~Logan and S.~Mrenna,
  Phys.\ Rev.\ D {\bf 65}, 055005 (2002), \\
  Erratum: [Phys.\ Rev.\ D {\bf 65}, 099902 (2002)]
  [hep-ph/0106116].

\bibitem{Kanemura:2017wtm} 
  S.~Kanemura, M.~Kikuchi, K.~Sakurai and K.~Yagyu,
  Phys.\ Rev.\ D {\bf 96}, no. 3, 035014 (2017)
  [arXiv:1705.05399 [hep-ph]].

\bibitem{Arhrib:2003ph} 
  A.~Arhrib, M.~Capdequi Peyranere, W.~Hollik and S.~Penaranda,
  Phys.\ Lett.\ B {\bf 579}, 361 (2004)
  [hep-ph/0307391].

\bibitem{Kanemura:2014dja} 
  S.~Kanemura, M.~Kikuchi and K.~Yagyu,
  Phys.\ Lett.\ B {\bf 731}, 27 (2014)
  [arXiv:1401.0515 [hep-ph]].

\bibitem{Kanemura:2015mxa} 
  S.~Kanemura, M.~Kikuchi and K.~Yagyu,
  Nucl.\ Phys.\ B {\bf 896}, 80 (2015)
  [arXiv:1502.07716 [hep-ph]].


\bibitem{LopezVal:2010vk}
  D.~Lopez-Val, J.~Sola and N.~Bernal,
  Phys.\ Rev.\ D {\bf 81}, 113005 (2010)
  [arXiv:1003.4312 [hep-ph]].


\bibitem{Altenkamp:2017ldc} 
  L.~Altenkamp, S.~Dittmaier and H.~Rzehak,
  arXiv:1704.02645 [hep-ph].


\bibitem{Kanemura:2015fra} 
  S.~Kanemura, M.~Kikuchi and K.~Yagyu,
  Nucl.\ Phys.\ B {\bf 907}, 286 (2016)
  [arXiv:1511.06211 [hep-ph]].

\bibitem{Kanemura:2016lkz} 
  S.~Kanemura, M.~Kikuchi and K.~Yagyu,
  Nucl.\ Phys.\ B {\bf 917}, 154 (2017)
  [arXiv:1608.01582 [hep-ph]].
  

\bibitem{He:2016sqr} 
  S.~P.~He and S.~h.~Zhu,
  Phys.\ Lett.\ B {\bf 764}, 31 (2017)
  [arXiv:1607.04497 [hep-ph]].



\bibitem{Kanemura:2016sos} 
  S.~Kanemura, M.~Kikuchi and K.~Sakurai,
  Phys.\ Rev.\ D {\bf 94}, no. 11, 115011 (2016)
  [arXiv:1605.08520 [hep-ph]].

\bibitem{Arhrib:2015hoa} 
  A.~Arhrib, R.~Benbrik, J.~El Falaki and A.~Jueid,
  JHEP {\bf 1512}, 007 (2015)
  [arXiv:1507.03630 [hep-ph]].

\bibitem{Djouadi:1997yw} 
  A.~Djouadi, J.~Kalinowski and M.~Spira,
  Comput.\ Phys.\ Commun.\  {\bf 108}, 56 (1998)
  [hep-ph/9704448].

\bibitem{Heinemeyer:1998yj} 
  S.~Heinemeyer, W.~Hollik and G.~Weiglein,
  Comput.\ Phys.\ Commun.\  {\bf 124}, 76 (2000)
  [hep-ph/9812320].
 
\bibitem{Hahn:2009zz} 
  T.~Hahn, S.~Heinemeyer, W.~Hollik, H.~Rzehak and G.~Weiglein,
  Comput.\ Phys.\ Commun.\  {\bf 180}, 1426 (2009).

\bibitem{Ellwanger:2004xm} 
  U.~Ellwanger, J.~F.~Gunion and C.~Hugonie,
  JHEP {\bf 0502}, 066 (2005)
  [hep-ph/0406215].


\bibitem{Eriksson:2009ws} 
  D.~Eriksson, J.~Rathsman and O.~Stal,
  Comput.\ Phys.\ Commun.\  {\bf 181}, 189 (2010)
  [arXiv:0902.0851 [hep-ph]].

\bibitem{Costa:2015llh} 
  R.~Costa, M.~M\"uhlleitner, M.~O.~P.~Sampaio and R.~Santos,
  JHEP {\bf 1606}, 034 (2016)
  [arXiv:1512.05355 [hep-ph]].


\bibitem{Yamada:2001px} 
  Y.~Yamada,
  Phys.\ Rev.\ D {\bf 64}, 036008 (2001)
  [hep-ph/0103046].

\bibitem{Espinosa:2002cd} 
  J.~R.~Espinosa and Y.~Yamada,
  Phys.\ Rev.\ D {\bf 67}, 036003 (2003)
  [hep-ph/0207351].




\bibitem{Kakizaki:2013eba} 
  M.~Kakizaki, S.~Kanemura, H.~Taniguchi and T.~Yamashita,
  Phys.\ Rev.\ D {\bf 89}, no. 7, 075013 (2014)
  [arXiv:1312.7575 [hep-ph]].

\bibitem{Kanemura:2014bqa} 
  S.~Kanemura, K.~Tsumura, K.~Yagyu and H.~Yokoya,
  Phys.\ Rev.\ D {\bf 90}, 075001 (2014)
  [arXiv:1406.3294 [hep-ph]].

\bibitem{Kanemura:2014kga} 
  S.~Kanemura, K.~Kaneta, N.~Machida and T.~Shindou,
  Phys.\ Rev.\ D {\bf 91}, 115016 (2015)
  [arXiv:1410.8413 [hep-ph]].

\bibitem{Fujii:2015jha} 
  K.~Fujii {\it et al.},
  arXiv:1506.05992 [hep-ex].

\bibitem{Chen:2014ask} 
  C.~Y.~Chen, S.~Dawson and I.~M.~Lewis,
  Phys.\ Rev.\ D {\bf 91}, no. 3, 035015 (2015)
  [arXiv:1410.5488 [hep-ph]].
 
\bibitem{Barger:1989fj} 
  V.~D.~Barger, J.~L.~Hewett and R.~J.~N.~Phillips,
  Phys.\ Rev.\ D {\bf 41}, 3421 (1990).

\bibitem{Grossman:1994jb} 
  Y.~Grossman,
  Nucl.\ Phys.\ B {\bf 426}, 355 (1994)
  [hep-ph/9401311].

\bibitem{Aoki:2009ha} 
  M.~Aoki, S.~Kanemura, K.~Tsumura and K.~Yagyu,
  Phys.\ Rev.\ D {\bf 80}, 015017 (2009)
  [arXiv:0902.4665 [hep-ph]].
 


\bibitem{Lee:1977eg} 
  B.~W.~Lee, C.~Quigg and H.~B.~Thacker,
  Phys.\ Rev.\ D {\bf 16}, 1519 (1977).
  
  


  
\bibitem{Cynolter:2004cq} 
  G.~Cynolter, E.~Lendvai and G.~Pocsik,
  Acta Phys.\ Polon.\ B {\bf 36}, 827 (2005)
  [hep-ph/0410102].
  
\bibitem{Kanemura:1993hm} 
  S.~Kanemura, T.~Kubota and E.~Takasugi,
  Phys.\ Lett.\ B {\bf 313}, 155 (1993)
  [hep-ph/9303263].
  
\bibitem{Akeroyd:2000wc} 
  A.~G.~Akeroyd, A.~Arhrib and E.~M.~Naimi,
  Phys.\ Lett.\ B {\bf 490}, 119 (2000)
  [hep-ph/0006035].
  
\bibitem{Ginzburg:2005dt} 
  I.~F.~Ginzburg and I.~P.~Ivanov,
  Phys.\ Rev.\ D {\bf 72}, 115010 (2005)
  [hep-ph/0508020].
  
\bibitem{Kanemura:2015ska} 
  S.~Kanemura and K.~Yagyu,
  Phys.\ Lett.\ B {\bf 751}, 289 (2015)
  [arXiv:1509.06060 [hep-ph]].

\bibitem{Gonderinger:2009jp} 
  M.~Gonderinger, Y.~Li, H.~Patel and M.~J.~Ramsey-Musolf,
  JHEP {\bf 1001}, 053 (2010)
  [arXiv:0910.3167 [hep-ph]].
  
\bibitem{Inoue:1982ej} 
  K.~Inoue, A.~Kakuto, H.~Komatsu and S.~Takeshita,
  Prog.\ Theor.\ Phys.\  {\bf 67}, 1889 (1982).
  
\bibitem{Goudelis:2013uca} 
  A.~Goudelis, B.~Herrmann and O.~St\aa l,
  JHEP {\bf 1309}, 106 (2013)
  [arXiv:1303.3010 [hep-ph]].
  
\bibitem{Pruna:2013bma} 
  G.~M.~Pruna and T.~Robens,
  Phys.\ Rev.\ D {\bf 88}, no. 11, 115012 (2013)
  [arXiv:1303.1150 [hep-ph]].
  
\bibitem{Fuyuto:2014yia} 
  K.~Fuyuto and E.~Senaha,
  Phys.\ Rev.\ D {\bf 90}, no. 1, 015015 (2014)
  [arXiv:1406.0433 [hep-ph]].
  
\bibitem{Robens:2015gla} 
  T.~Robens and T.~Stefaniak,
  Eur.\ Phys.\ J.\ C {\bf 75}, 104 (2015)
  [arXiv:1501.02234 [hep-ph]].
  
\bibitem{Deshpande:1977rw} 
  N.~G.~Deshpande and E.~Ma,
  Phys.\ Rev.\ D {\bf 18}, 2574 (1978).
 
\bibitem{Klimenko:1984qx} 
  K.~G.~Klimenko,
  Theor.\ Math.\ Phys.\  {\bf 62}, 58 (1985)
  [Teor.\ Mat.\ Fiz.\  {\bf 62}, 87 (1985)].
  
\bibitem{Sher:1988mj} 
  M.~Sher,
  Phys.\ Rept.\  {\bf 179}, 273 (1989);
  
  S.~Nie and M.~Sher,
  Phys.\ Lett.\ B {\bf 449}, 89 (1999)
  [hep-ph/9811234].
  
\bibitem{Kanemura:1999xf} 
  S.~Kanemura, T.~Kasai and Y.~Okada,
  Phys.\ Lett.\ B {\bf 471}, 182 (1999)
  [hep-ph/9903289].
  
\bibitem{Espinosa:2011ax} 
  J.~R.~Espinosa, T.~Konstandin and F.~Riva,
  Nucl.\ Phys.\ B {\bf 854}, 592 (2012)
  [arXiv:1107.5441 [hep-ph]].
  

\bibitem{Peskin:1990zt} 
  M.~E.~Peskin and T.~Takeuchi,
  Phys.\ Rev.\ Lett.\  {\bf 65}, 964 (1990).
  
  M.~E.~Peskin and T.~Takeuchi,
  Phys.\ Rev.\ D {\bf 46}, 381 (1992).

\bibitem{Baak:2012kk} 
  M.~Baak {\it et al.},
  Eur.\ Phys.\ J.\ C {\bf 72}, 2205 (2012)
  [arXiv:1209.2716 [hep-ph]].
  
\bibitem{Lopez-Val:2014jva} 
  D.~Lopez-Val and T.~Robens,
  Phys.\ Rev.\ D {\bf 90}, 114018 (2014)
  [arXiv:1406.1043 [hep-ph]].
  
\bibitem{Toussaint:1978zm} 
  D.~Toussaint,
  Phys.\ Rev.\ D {\bf 18}, 1626 (1978).
  
\bibitem{Bertolini:1985ia} 
  S.~Bertolini,
  Nucl.\ Phys.\ B {\bf 272}, 77 (1986).
  
\bibitem{Peskin:2001rw} 
  M.~E.~Peskin and J.~D.~Wells,
  Phys.\ Rev.\ D {\bf 64}, 093003 (2001)
  [hep-ph/0101342].
  
\bibitem{Grimus:2008nb} 
  W.~Grimus, L.~Lavoura, O.~M.~Ogreid and P.~Osland,
  Nucl.\ Phys.\ B {\bf 801}, 81 (2008)
  [arXiv:0802.4353 [hep-ph]].
  
\bibitem{Kanemura:2011sj} 
  S.~Kanemura, Y.~Okada, H.~Taniguchi and K.~Tsumura,
  Phys.\ Lett.\ B {\bf 704}, 303 (2011)
  [arXiv:1108.3297 [hep-ph]].
  


 

\bibitem{Hahn} 
  T.~Hahn and M.~Perez-Victoria,
  Comput.\ Phys.\ Commun.\  {\bf 118}, 153 (1999)
  [hep-ph/9807565].

\bibitem{hqq_alp4} 
  P.~A.~Baikov, K.~G.~Chetyrkin and J.~H.~Kuhn,
  Phys.\ Rev.\ Lett.\  {\bf 96}, 012003 (2006)
  [hep-ph/0511063].

\bibitem{hqq_alpmix} 
  L.~Mihaila, B.~Schmidt and M.~Steinhauser,
  Phys.\ Lett.\ B {\bf 751}, 442 (2015)
  [arXiv:1509.02294 [hep-ph]].










\bibitem{hgg_nlo_1} 
  A.~Djouadi, M.~Spira and P.~M.~Zerwas,
  Phys.\ Lett.\ B {\bf 264}, 440 (1991).

\bibitem{hgg_nlo_2} 
  S.~Dawson,
  Nucl.\ Phys.\ B {\bf 359}, 283 (1991).

\bibitem{hgg_nlo_3} 
  M.~Spira, A.~Djouadi, D.~Graudenz and P.~M.~Zerwas,
  Nucl.\ Phys.\ B {\bf 453}, 17 (1995)
  [hep-ph/9504378].


\bibitem{djouadi} 
  A.~Djouadi,
  Phys.\ Rept.\  {\bf 457}, 1 (2008)
  [hep-ph/0503172].


\bibitem{hgg_nnlo_1} 
  R.~V.~Harlander and W.~B.~Kilgore,
  Phys.\ Rev.\ Lett.\  {\bf 88}, 201801 (2002)
  [hep-ph/0201206].

\bibitem{hgg_nnlo_2} 
  C.~Anastasiou and K.~Melnikov,
  Nucl.\ Phys.\ B {\bf 646}, 220 (2002)
  [hep-ph/0207004].

\bibitem{hgg_nnlo_3} 
  V.~Ravindran, J.~Smith and W.~L.~van Neerven,
  Nucl.\ Phys.\ B {\bf 665}, 325 (2003)
  [hep-ph/0302135].


\bibitem{hgg_nnlo_mt_1}
  S.~Marzani, R.~D.~Ball, V.~Del Duca, S.~Forte and A.~Vicini,
  Nucl.\ Phys.\ B {\bf 800} (2008) 127
  [arXiv:0801.2544 [hep-ph]].

\bibitem{hgg_nnlo_mt_2} 
  R.~V.~Harlander and K.~J.~Ozeren,
  Phys.\ Lett.\ B {\bf 679}, 467 (2009)  [arXiv:0907.2997 [hep-ph]]; 
  JHEP {\bf 0911}, 088 (2009)  [arXiv:0909.3420 [hep-ph]].

\bibitem{hgg_nnlo_lhc} 
  C.~Anastasiou, S.~Buehler, F.~Herzog and A.~Lazopoulos,
  JHEP {\bf 1204}, 004 (2012)
  [arXiv:1202.3638 [hep-ph]].


\bibitem{hgg_n3lo} 
  C.~Anastasiou, C.~Duhr, F.~Dulat, F.~Herzog and B.~Mistlberger,
  Phys.\ Rev.\ Lett.\  {\bf 114}, 212001 (2015)
  [arXiv:1503.06056 [hep-ph]].





\end{thebibliography}
\end{document}